\documentclass{elsart}
\usepackage{graphicx}
\usepackage{amssymb}

\begin{document}

\begin{frontmatter}


\title{Time asymmetries in extensive air showers: a novel method to identify UHECR species}
%
\author[label1]{M. T. Dova, M. E. Mance\~nido, A. G. Mariazzi, H. Wahlberg}
\address[label1]{IFLP (CONICET - La Plata/Universidad Nacional de La Plata),CC67 - 1900, La Plata, Argentina}
\author[label2]{F. Arqueros, D. Garc{{\'i}}a-Pinto}
\address[label2]{Departamento de F{{\'i}}sica At\'omica, Molecular y Nuclear. Facultad de Ciencias F{{\'i}}sicas, Universidad Complutense de Madrid, Madrid E-28040, Spain}

\begin{abstract}
Azimuthal asymmetries in signals of non vertical showers have been observed 
in ground arrays of water Cherenkov detectors, like Haverah Park and the Pierre Auger Observatory. 
The asymmetry in time distributions of arriving particles 
offers a new possibility for the determination of the mass composition.
The dependence of this asymmetry on atmospheric depth 
shows a clear maximum at a position that is correlated with the primary species. 
In this work a novel method to determine mass composition based on these features
of the ground signals is presented and a Monte Carlo study of its sensitivity is carried out.

\end{abstract}


\end{frontmatter}

\section{Introduction}
\label{sec:intro}
The determination of the nature of ultra high energy cosmic rays (UHECR) is a crucial
point to help understanding their origin, acceleration mechanisms and propagation from the sources to the Earth. 
At energies below $10^{15}$ {\rm eV}, both charge and mass can be measured 
directly 
using space detectors, however, the properties of cosmic rays of the highest energies 
have to be inferred from the features of the shower induced in the atmosphere. 
Air shower experiments are either ground arrays of detectors that trigger in coincidence
when a shower passes through them, or optical detectors that observe the longitudinal
development of the extensive air shower (EAS) 
\cite{Yoshida:1998jr,Linsley:prl63,HPInst,Baltrusaitis:ce,Abu-Zayyad:uu,Abraham}. 

The measurement of the primary mass in EAS experiments is known to be
very difficult due to the large fluctuations resulting from the statistical nature of the 
shower development, in particular those associated to the depth and the number of particles 
produced in the first interactions. 
Furthermore, the interpretation of data to determine mass composition has to be obtained 
by comparison with Monte Carlo predictions dependent on high energy hadronic models.
With increasing primary energy, this task becomes more difficult 
as the gap to the energy range studied in accelerator experiments
increases and the hadronic interaction properties have to be extrapolated over a wide range. 
One of the main sources of uncertainties in any analysis to determine mass composition comes 
from the different predictions for different hadronic interaction models. 

\begin{figure}[htb]
  \centering
 \includegraphics[width=14cm]{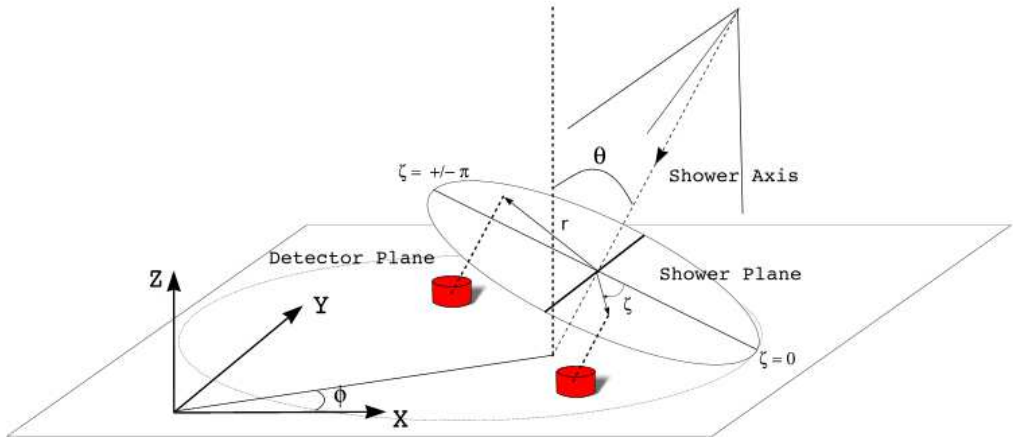} 
  \caption{Schematic view of the shower geometry. 
The vertical projection of the incoming direction into the shower plane ($\zeta = 0$) defines two regions, ``early'' ($\frac{-\pi}{2}< \zeta <\frac{\pi}{2}$)  before the shower core impact point 
  and the opposite ``late'' region. Note the different amount of atmosphere traversed by the particles reaching the detectors in each region.}
  \label{fig:esquema}
\end{figure}

The distribution of shower maximum, $X_{\rm{max}}$, that is the atmospheric depth at which the 
number of charged particles in the EAS is maximum, is sensitive to the composition of cosmic rays.
Protons produce deeper showers with fluctuations larger than those of heavier nuclei.
Therefore, for a given primary energy, the $<X_{\rm{max}}>$ value and its fluctuations 
decrease with heavier primary mass. This is the principle in which separation methods using  
$<X_{\rm{max}}>$, its fluctuations and the elongation rate, d($X_{\rm{max}}$)/d({\rm log}$E$)
\cite{ER:Linsley1977b, ER:AlvarezMuniz}, as measured by fluorescence detectors, are based.

In ground array experiments the analysis is usually performed by projecting the signals 
registered by the detectors into the shower plane (see Figure \ref{fig:esquema}) 
and thus, neglecting the further shower evolution of the late regions. 
As a consequence, for inclined showers, the circular symmetry in the signals of surface detectors 
is broken. This results in a dependence of the signal 
features on the azimuth angle in the shower plane ~\cite{Dova:2003rz}, mainly due to the different amount 
of atmosphere traversed by the shower particles ~\cite{Dova:2001jy}.

Evidences of azimuthal asymmetries in the signal size were first observed 
at Haverah Park~\cite{Engl}. Recently the Pierre Auger Observatory has found in addition, 
the expected asymmetry in the particle arrival time distributions ~\cite{Dova:2003rz}.
The observation of these asymmetries for incoming directions with zenith angle smaller than 60$^{\circ}$, 
has been possible at the Pierre Auger Observatory due to the large size of the array and the high time
resolution electronic of the surface detector stations ~\cite{Nitz}. 
The design of the observatory allows measuring this feature of EAS which, as demonstrated below, 
carries very valuable information on to the chemical composition of cosmic rays. 
First results showing the sensitivity to primary species at the Pierre Auger Observatory
have been presented in ~\cite{Dova:2005pune}  and ~\cite{Healy}.

In this work it is shown that the asymmetry in risetime, $t_{1/2}$, defined as the time to reach 
from 10\% to 50\% of the total integrated signal in each station, 
is related to the shower stage of development ~\cite{Healy,HP,HaverahPark}. 
Thus, for a given primary energy $E$, the asymmetry depends on zenith angle $\theta$
of the primary cosmic ray in such a way that its behavior versus $\sec \theta$  
is reminiscent of the longitudinal development of the shower. 
This ``longitudinal development of the asymmetry'' is strongly dependent on the nature of the primary particle. 
The method presented here is quite general and, in principle, might be applied to other timing parameters 
describing the time signal structure as well as other shower observables like signal size which was observed 
to be less sensitive to mass composition.

The analysis described in this work is based on Monte Carlo simulations carried out with the code 
{\sc aires} ~\cite{Aires} using the hadronic interaction models {\sc qgsjetii}(03) \cite{qgsjetII} 
and {\sc sibyll} 2.1 \cite{sibyll21}.
The {\sc aires} generated showers were subsequently used as input 
in the detector simulation code and finally reconstructed using for both tasks 
the official Offline reconstruction framework of the Pierre Auger Observatory ~\cite{DPA}.

The paper is organised as follows. 
In section \ref{sec:linear} the relationship between asymmetry in the time structure 
and shower evolution is discussed in detail. 
A brief description of the Pierre Auger Observatory, which has been selected as a case study 
of this novel technique is given in section \ref{sec:au}.
The features of the Monte Carlo sample used for our analysis are described in section \ref{sec:mc}. 
The core of the method is presented in section \ref{sec:steps} where several mass sensitive parameters are defined. 
The procedure for the determination of the longitudinal asymmetry development and the 
definition of the discrimination parameters is firstly presented in \ref{sec:longdev}. 
The energy dependence of the parameters as predicted by two different hadronic models 
is discussed in \ref{sec:had}. In section \ref{sec:xmax} 
the relationship of our discrimination parameters with the shower maximum depth is analysed.
Finally, a statistical method to estimate the primary mass composition under a 
two-component (p-Fe) assumption is presented in section \ref{sec:fefraction} 
with a discussion on systematic uncertainties.  

\section{Asymmetry in the time structure as an indication of shower evolution}
\label{sec:linear}
Most of the observables sensitive to composition aim at being like a snapshot of the shower development
and thus, they are correlated with both $X_{\rm{max}}$ and the observation depth.
The time distribution of the signals recorded by the surface detector in EAS experiments contains
 implicitly the information of the shower development. Therefore, it is natural to expect a dependence of 
the mean value of risetime and its azimuthal asymmetry with the atmospheric depth traversed 
by the shower and thus with the zenith angle of the incoming cosmic ray direction.

\begin{figure}[htb]
  \centering
 \includegraphics[width=14cm]{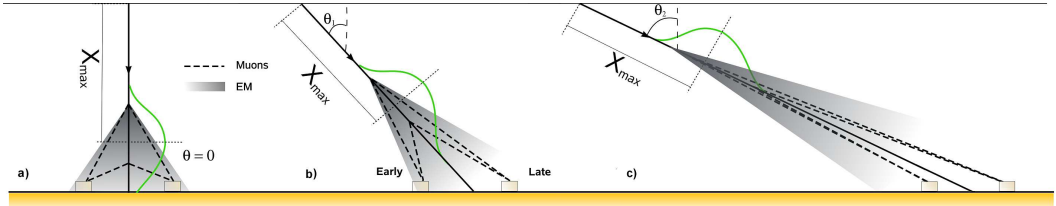} 
  \caption{Schematic view of shower development when arriving at three different zenith angles}
  \label{fig:eas}
\end{figure}

The sensitivity of timing parameters to primary composition can be explained on the basis of 
the dominance in the different time regions of the signal of the electromagnetic and muonic component.
The first portion of the signal is dominated by the muon component which tends 
to arrive earlier and over a period of time shorter than that of the electromagnetic particles (EM), 
which are spread out on time.

The relation between asymmetry and shower evolution is sketched in Figure ~\ref{fig:eas}, 
where three different scenarios are presented for a shower with a given $X_{\rm{max}}$ value,
arriving at three different zenith angles.
The attenuation (early-late) of the EM component depends on the difference 
in the path travelled by particles to reach the detector stations. 
In case (a), i.e. vertical shower, there is no difference in the paths of the EM component 
so there is no asymmetry. As the zenith angle increases (case b) 
the difference in the attenuation of the EM component due to different travelled paths 
increases and, as a result, early-late asymmetry appears. At zenith angles smaller 
than 30$^0$ there is an additional effect affecting the asymmetry when using water Cherenkov 
detector for surface arrays, due to a combination of the geometry of these stations and the 
arrival direction of individual particles \cite{Dova:2005pune,Bertou}. At very large zenith 
angles (case c) the EM component has been strongly absorbed before 
reaching the detector. Therefore, the  asymmetry decreases with $\theta$ since the larger 
is the angle the smaller is the contribution of the EM component. 
Note that the muonic component is basically asymmetry free. 
Then for a given zenith angle the asymmetry gives information of the 
stage of development of the shower. According to the above arguments, a plot of 
asymmetry against ${\rm sec} \,\theta$ is expected to have a maximum 
which is correlated with the longitudinal shower evolution. 

The slant depth traversed by the shower particles,  $t'$, can be expressed using a Taylor expansion 
around $t_s = t\,{\rm sec} \theta$, where $t$ is the vertical atmospheric
depth at ground. Since the azimuth angle correction is small compared to the total slant depth,
in a first approximation, one can keep only the first term which is equivalent
to using a linear function in  ${\rm cos}\, \zeta$ to describe asymmetries \cite{Dova:2001jy}.
Thus, the atmospheric slant depth around the shower axis, can be expressed at fixed distance 
from the core, $r=r_0$, by
\begin{equation}
t'(\zeta, r=r_0) =  t\,\sec\theta\,(1+B\,{\rm cos}\,\zeta) = t_s\,+\,\Delta t_s(\zeta)
\end{equation}
A generic time distribution for vertical showers $\tau(r,t)$ depending on 
atmospheric depth and core distance $r$, becomes for inclined showers 
\begin{equation}
\tau(r,t)\, \,\rightarrow\, \,\tau(r,t'(\zeta,\theta))
\end{equation}
where $r$ is measured in the shower plane.

A Taylor expansion of $\tau$ around $t_s$ gives
\begin{equation}
\tau(r,\zeta) = \tau(r,t'(\zeta))=\tau(r,t_s+\Delta t_s(\zeta))
\end{equation}
\begin{equation}
\tau(r,\zeta) = \tau(r,t_s)+\frac{\partial\tau}{\partial t'}|_{t_s}\,\Delta t_s(\zeta)+...
\end{equation}
\begin{equation}
\tau(r,\zeta) = \tau(r,t_s)+\frac{\partial\tau}{\partial t'}|_{t_s}\,\,t_s\,B\,\cos\zeta+...
\end{equation}
\begin{equation}
\tau(r,\zeta) = \tau(r,t_s)(1+\frac{\partial{\rm ln}\tau}{\partial{\rm ln} t'}|_{t_s}\, B\,\cos\zeta+...)  \label{eq:taylor}
\end{equation}

Keeping only the first term of the expansion, equation (\ref{eq:taylor}) can be expressed as
\begin{equation}
\tau(r,\zeta) =  a + b\,{\rm cos}\, \zeta \label{eq:linearcos}
\end{equation}
with
\begin{equation}
a = \tau(r,t \,{\rm sec}\,\theta)
\end{equation}
\begin{equation}
\frac{b}{a} = B\,\frac{\partial{\rm ln}\tau}{\partial{\rm ln} t'}|_{t_s}
\end{equation}
The asymmetry factor $b/a$ which depends on $t_s$,
can be used as a measure of the logarithmic variation of $\tau$ with slant depth.
This parameter is an indicator of the shower evolution and hence it provides a measure 
of the composition of the primary particle. 
The dependence of the asymmetry factor $\frac{b}{a}$ with ${\rm sec} \theta$ allows one 
to find new observables useful for determining the mass composition, as it will 
be shown in Section \ref{sec:steps}.

\section{The Pierre Auger Observatory}
\label{sec:au}
The technique shown in this work relies in the fact that ground detectors located symmetrically around the shower axis are spread out over large areas. Therefore, the EM component of the late signal is attenuated as compared with that of the early one. This attenuation is only significant if the shower front is very extensive. The smaller is the primary energy, the shorter is the difference in atmospheric depth between early and late stations and thus, the weaker is the observed asymmetry. For instance, a reduction of the energy shower from $10^{18}$ to $10^{17}$ {\rm eV} lowers the maximum distance between fired detectors by about a 50\% and so the same percentage in the attenuation difference. 
On the other hand, the asymmetry in the time structure of the signal can be observed if the detectors are sensitive to both the EM and muonic components. Notice that the attenuation of the EM component translates into a decrease of the signal risetime because the earlier muon contribution is basically the same in the early and late regions (see Figure 2). Thus, this method can be applied in air-shower arrays designed for the detection of UHECR which are able to register information from both components.
The recently completed southern site of the Pierre Auger Observatory \cite{Abraham} fulfils these requirements and therefore the technique is very suitable for studies of mass composition using its ground array. The northern site planned to be built in Colorado USA, will also be able to use this technique very efficiently with a larger array. The two sites will ensure full sky coverage. Both have been designed to use hybrid detection (a ground array of Cherenkov detectors overlooked by fluorescence telescopes) to record EASs produced by cosmic rays of energies greater than $10^{18}$ eV. The Surface Detector (SD) of the southern site, situated near the town of 
Malarg\"ue, in western Argentina, consists of 1600 stations equally spaced on a triangular grid (1.5 {\rm km}) over an area of approximately 3000 ${\rm km}^2$. Each SD station is a water Cherenkov tank, 1.2 {\rm m} high and top area of 10 ${\rm m}^2$ with an internal reflecting coating. 
Three 9 inches PMTs overlook the water, and their signals are recorded by local digitisation electronics with a 40 {\rm MHz} sampling rate. The southern site Fluorescence Detector (FD) consists of 4 eyes with 6 telescopes each located at the border of the SD array overlooking it. 

The SD records the shower front, by sampling the particle density at ground level, with a duty cycle of 100\%. The FD measures the fluorescence light emitted as the shower develops through the atmosphere. As it can only operate on clear, moonless nights, its duty cycle is about 10\%. This unique combination of both techniques in a hybrid detector offers huge advantages in particular, for the determining energy \cite{Abraham:2008ru}. With respect to primary mass estimation, hybrid events provide a direct measurement of $X_{\rm max}$.
However, the bulk of events collected by the Observatory have information only from the surface array, making SD observables as the one presented in this paper very valuable for composition analysis at the highest energies.

\section{ Monte Carlo simulated data }
\label{sec:mc}
The sensitivity of the method proposed here for mass composition measurements 
has been studied, as mentioned above, with simulated showers generated with
{\sc aires} 2.8.3 ~\cite{Aires} using both hadronic 
interaction models {\sc qgsjetii}(03) and {\sc sibyll} 2.1.
 
The generated data sample contains a total of 2$\times 10^4$ showers, initiated
by proton and iron nuclei.
The energy and zenith angle values are as follows:
\begin{itemize}
\item{${\rm log}_{10}(E/{\rm eV})$ = 18.5, 19.0, 19.25, 19.5, 19.75, 20.0 } 
\item{$\theta$ (deg) =  32, 36, 41, 45, 49, 53, 57, 60, 63 }
\end{itemize}
The {\sc aires} generated showers were subsequently used as input in the detector simulation code 
and finally reconstructed using for both tasks the official Offline reconstruction framework 
of the Pierre Auger Observatory ~\cite{DPA}.
Detector simulation includes the generation of ground signals in the water Cherenkov stations 
of the Pierre Auger Observatory. The surface detector simulation has been tested and proved to be in good agreement with experimental data ~\cite{tanksim}.

The reconstructed values of the primary energy, arrival direction and core location of the shower 
have been used in this analysis. 
The analysis is limited to the range of reconstructed $\theta$ value above 30$^o$ and below zenith angle of 63$^o$ where the asymmetry effect due to the shower evolution dominates. In addition standard fiducial cuts on SD stations have been applied including minimum and maximum distances to the core, signal sizes and good reconstruction of global shower parameters.

\section{Mass sensitive parameters}
\label{sec:steps}

In this section several parameters related to the longitudinal asymmetry development 
will be defined and their sensitivity to the primary mass analysed.

Due to the intrinsic fluctuations in extensive air showers and the limited sampling of the shower front 
recorded by the SD, it is not possible to obtain the mass composition in a shower by shower basis. 
That is, instead of measuring the asymmetry in individual showers, the approach used in this work 
consists of using the mean value of the asymmetry factor, as defined above (equation \ref{eq:linearcos}), 
for all showers in a certain interval of energy and zenith angle.

\subsection{The longitudinal asymmetry development}
\label{sec:longdev}

\begin{figure}[htpb]
\includegraphics[width=7cm]{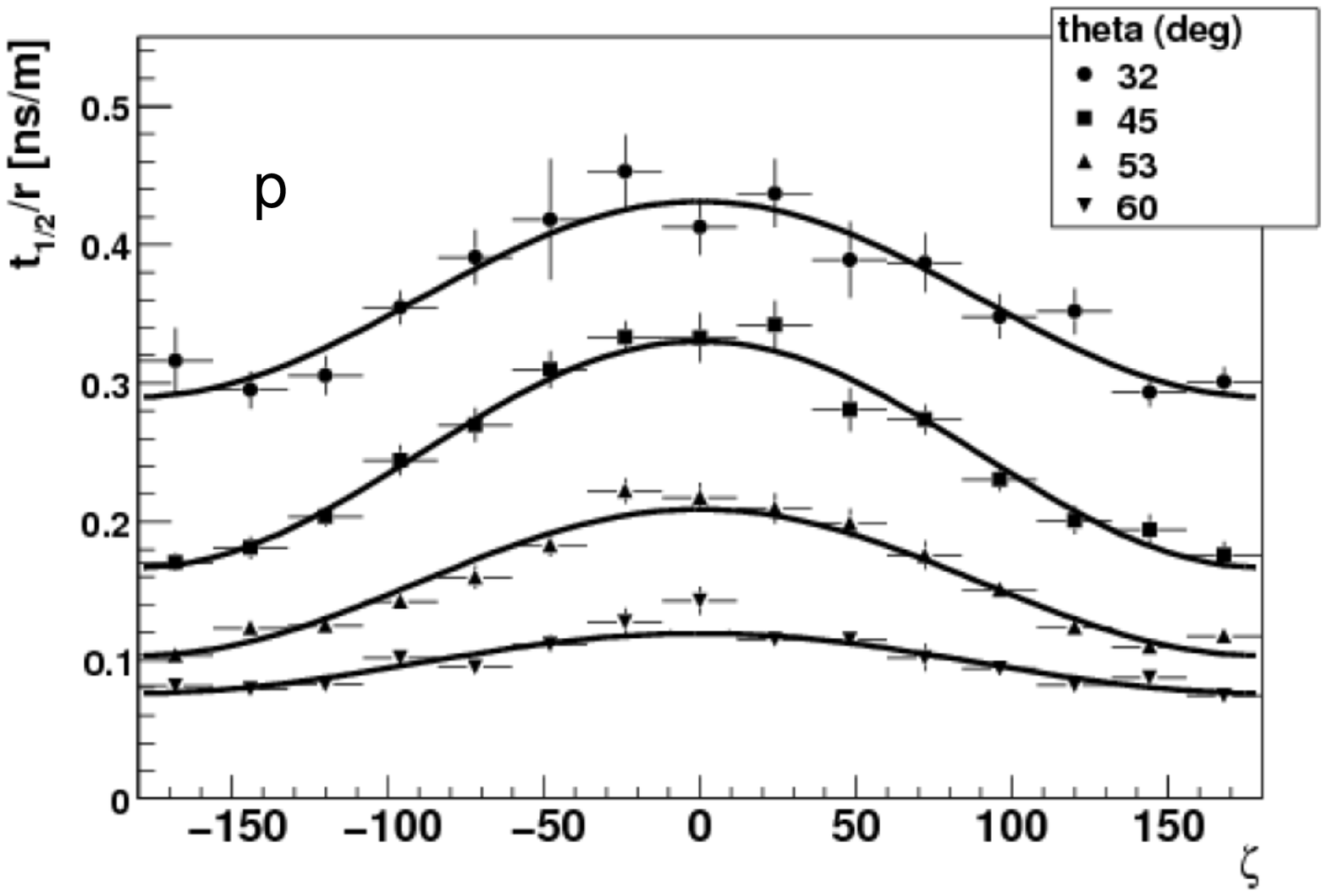}
\includegraphics[width=7cm]{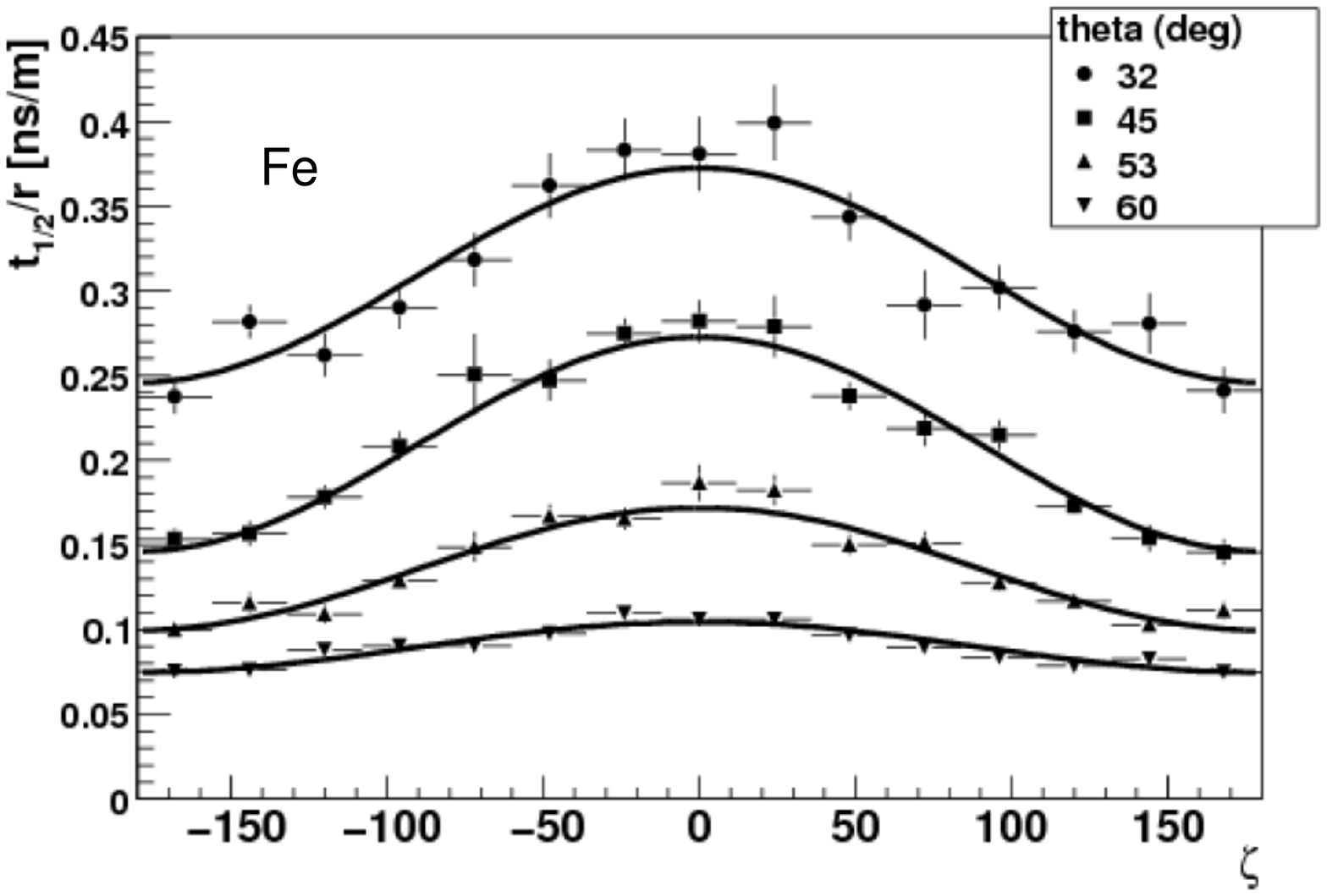}
\caption{Risetime versus azimuth angle in the shower plane for primary energy of $10^{19}$ {\rm eV} and different zenith angles. Proton (left) and iron (right).}
\label{fig:rise_vs_zeta_mc}
\end{figure}

\begin{figure}[htpb]
\includegraphics[width=7.cm]{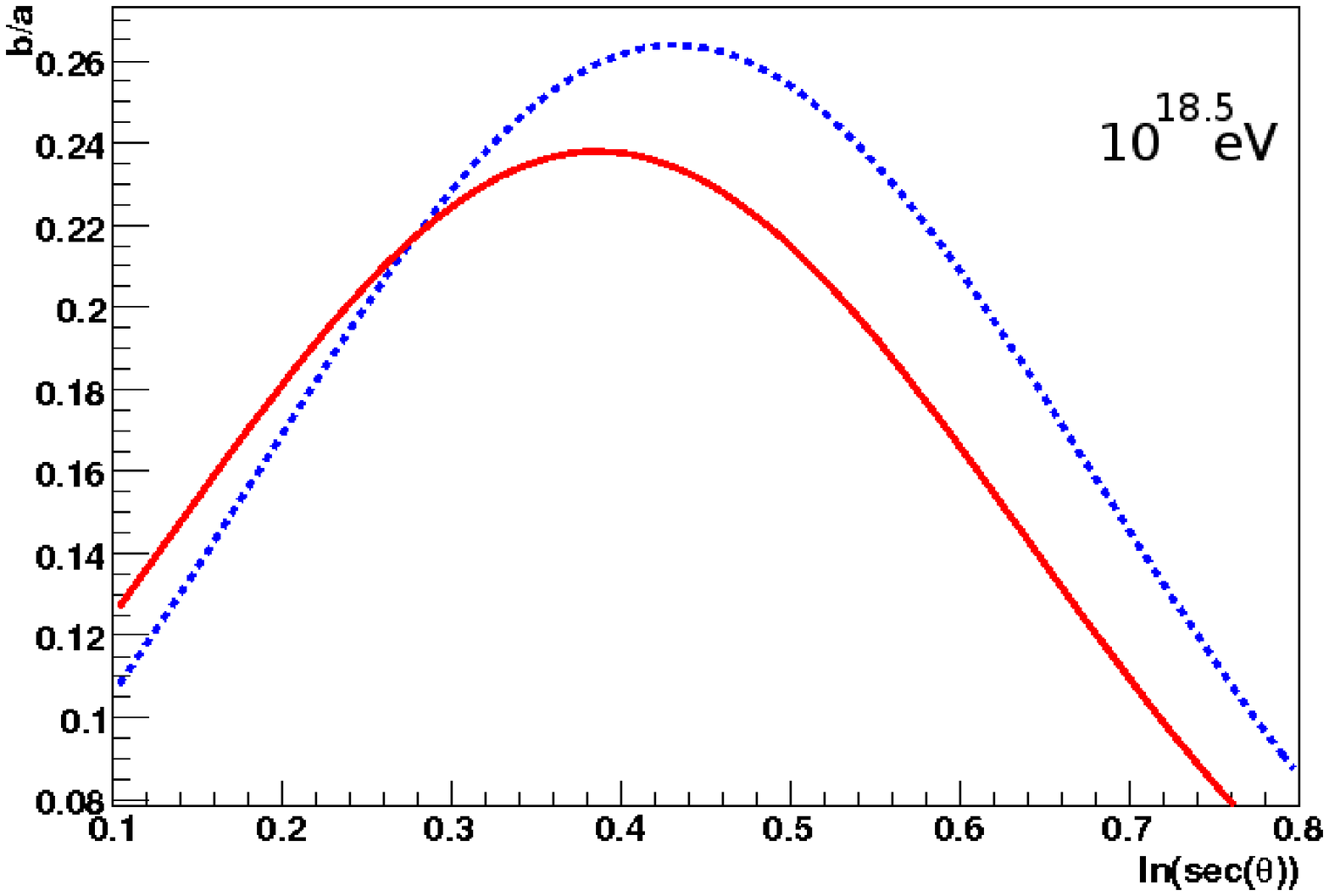}
\includegraphics[width=7.cm]{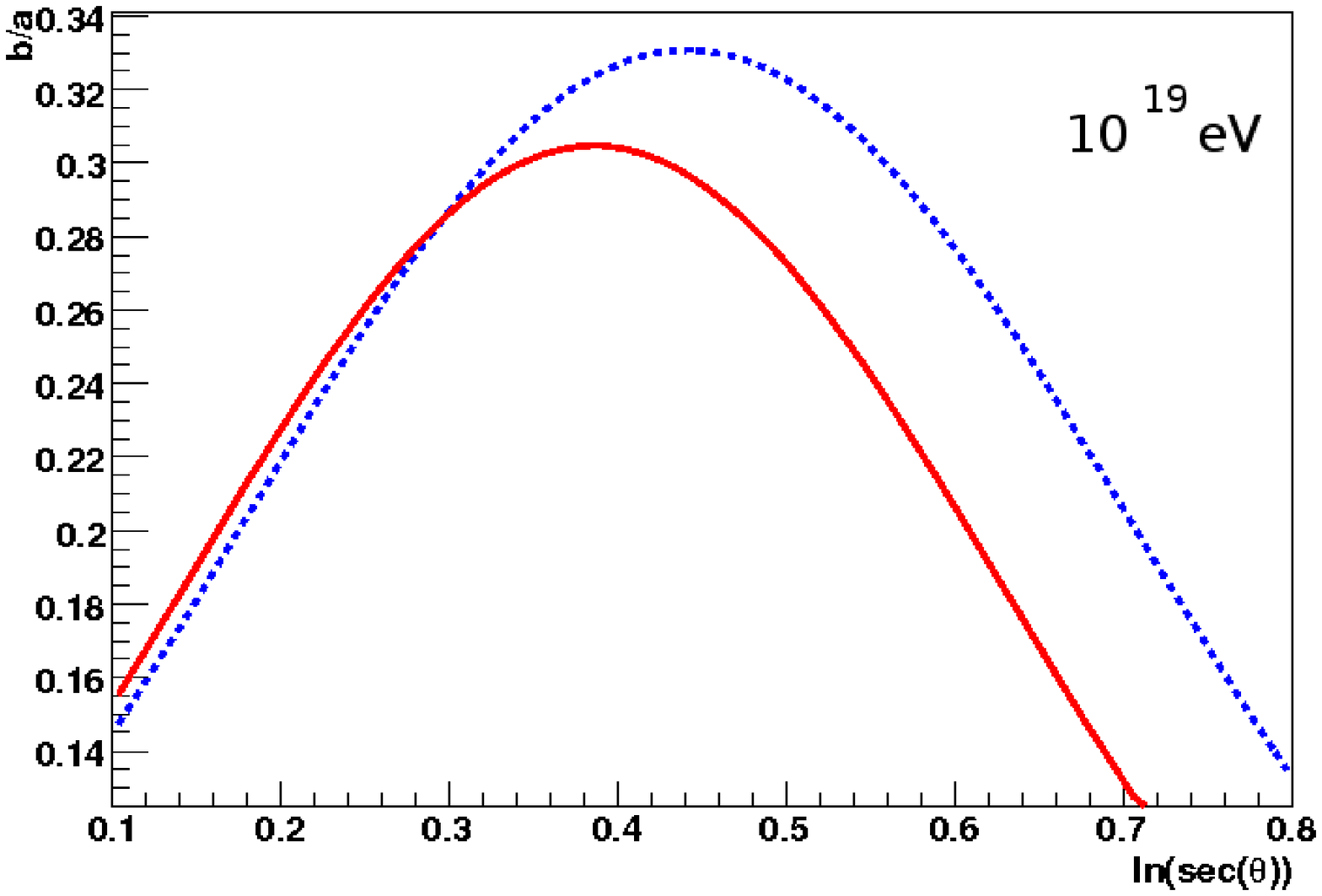}
\includegraphics[width=7.cm]{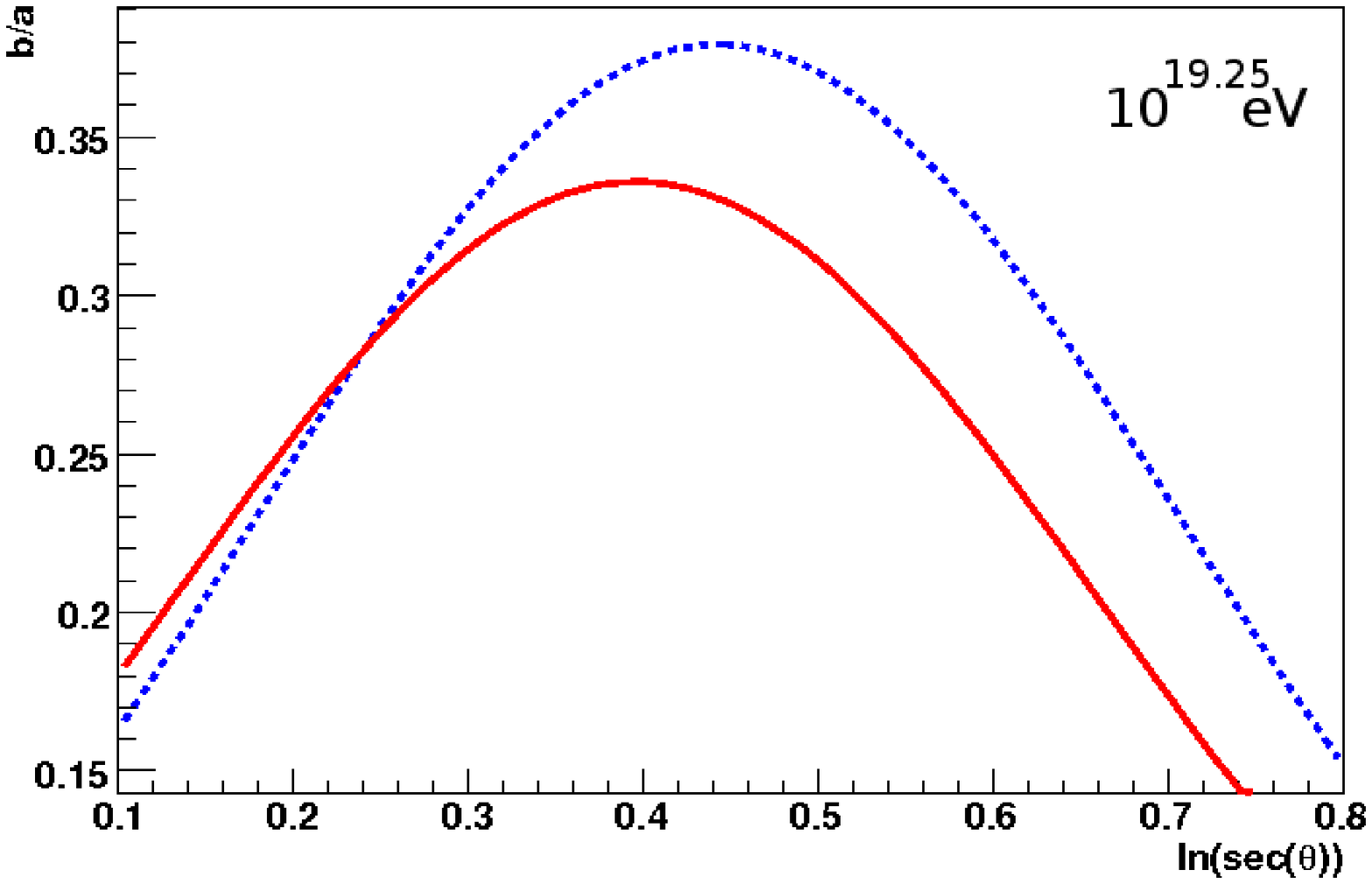}
\includegraphics[width=7.cm]{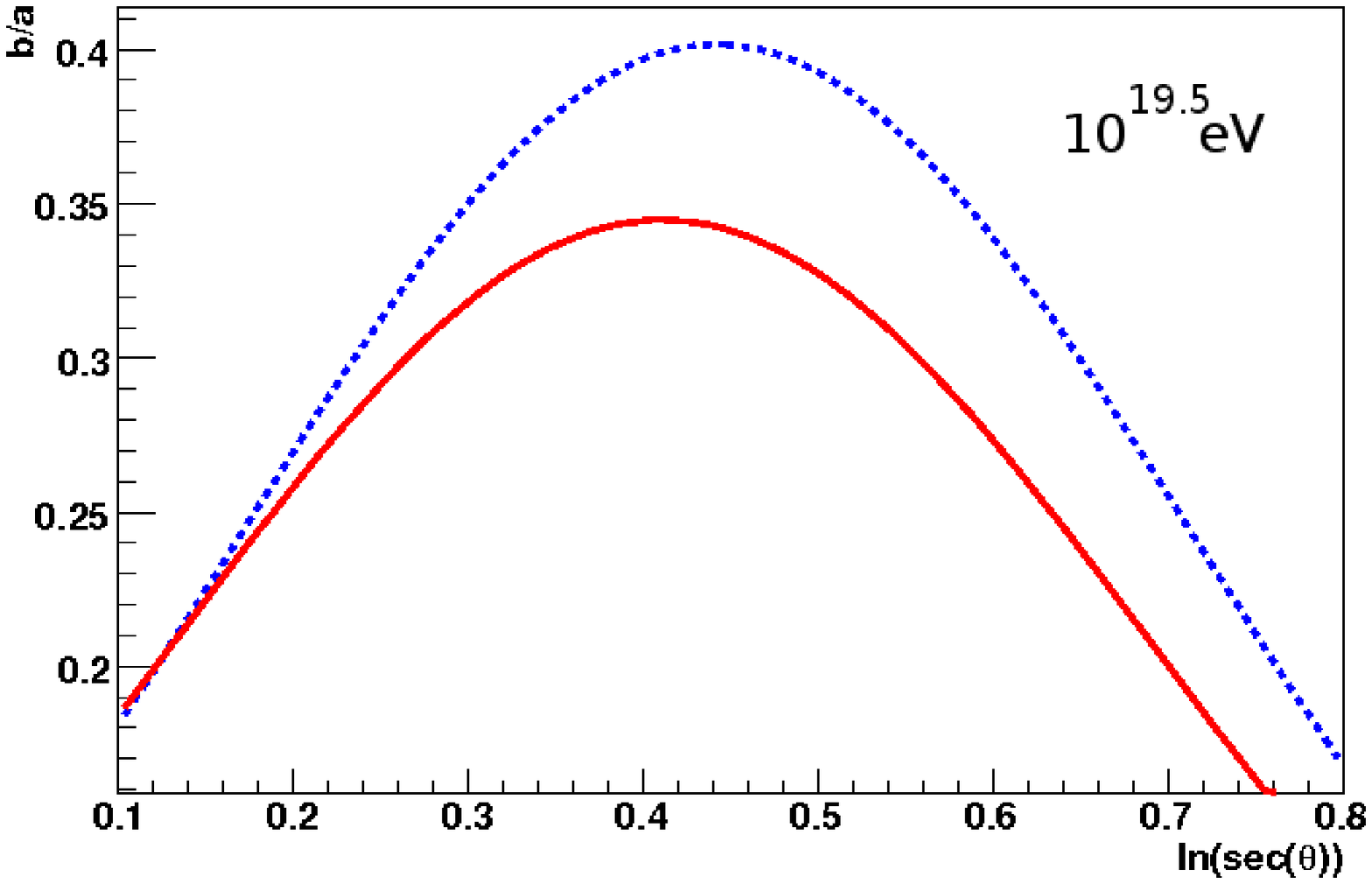}
\includegraphics[width=7.cm]{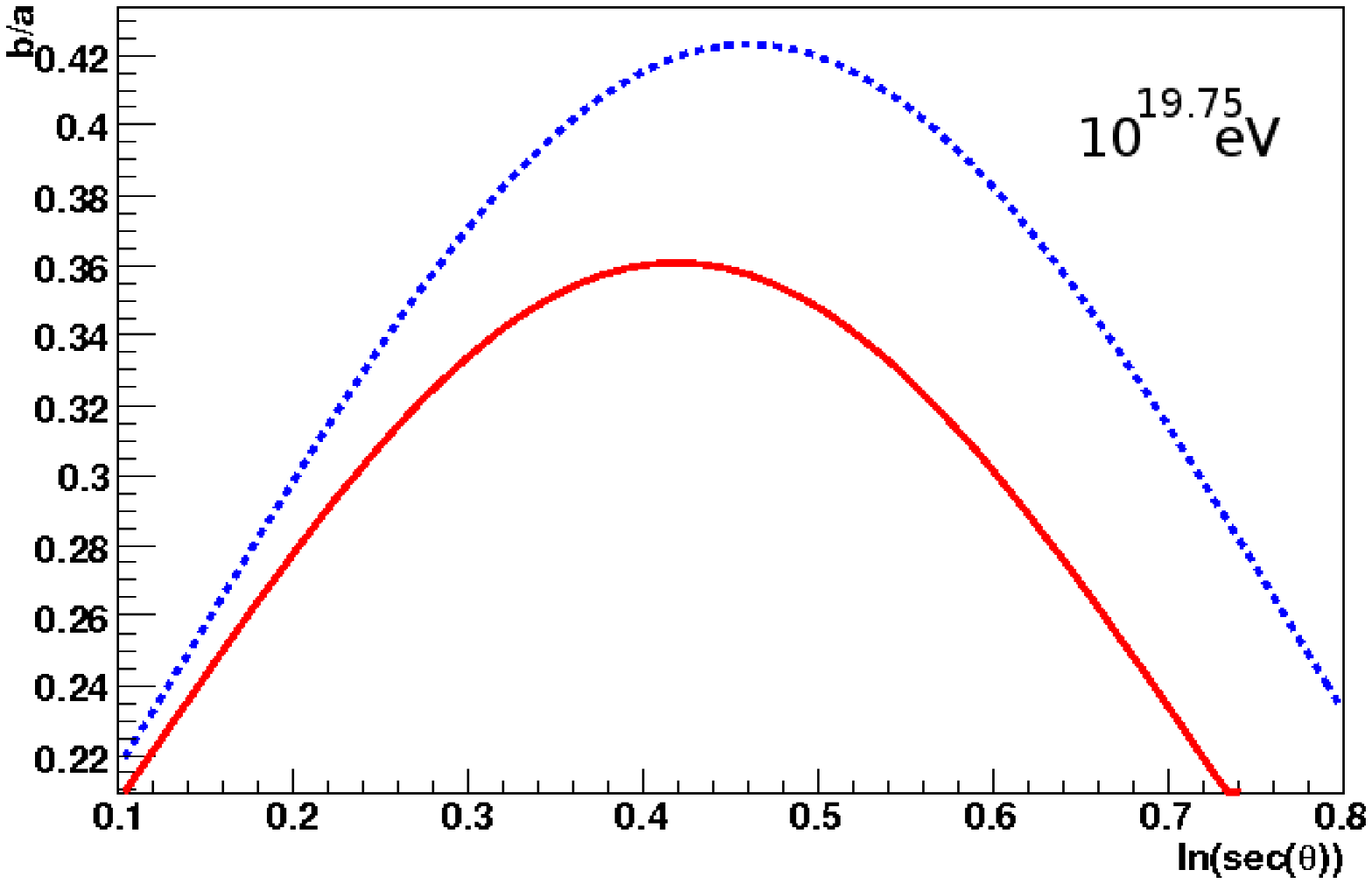}
\includegraphics[width=7.cm]{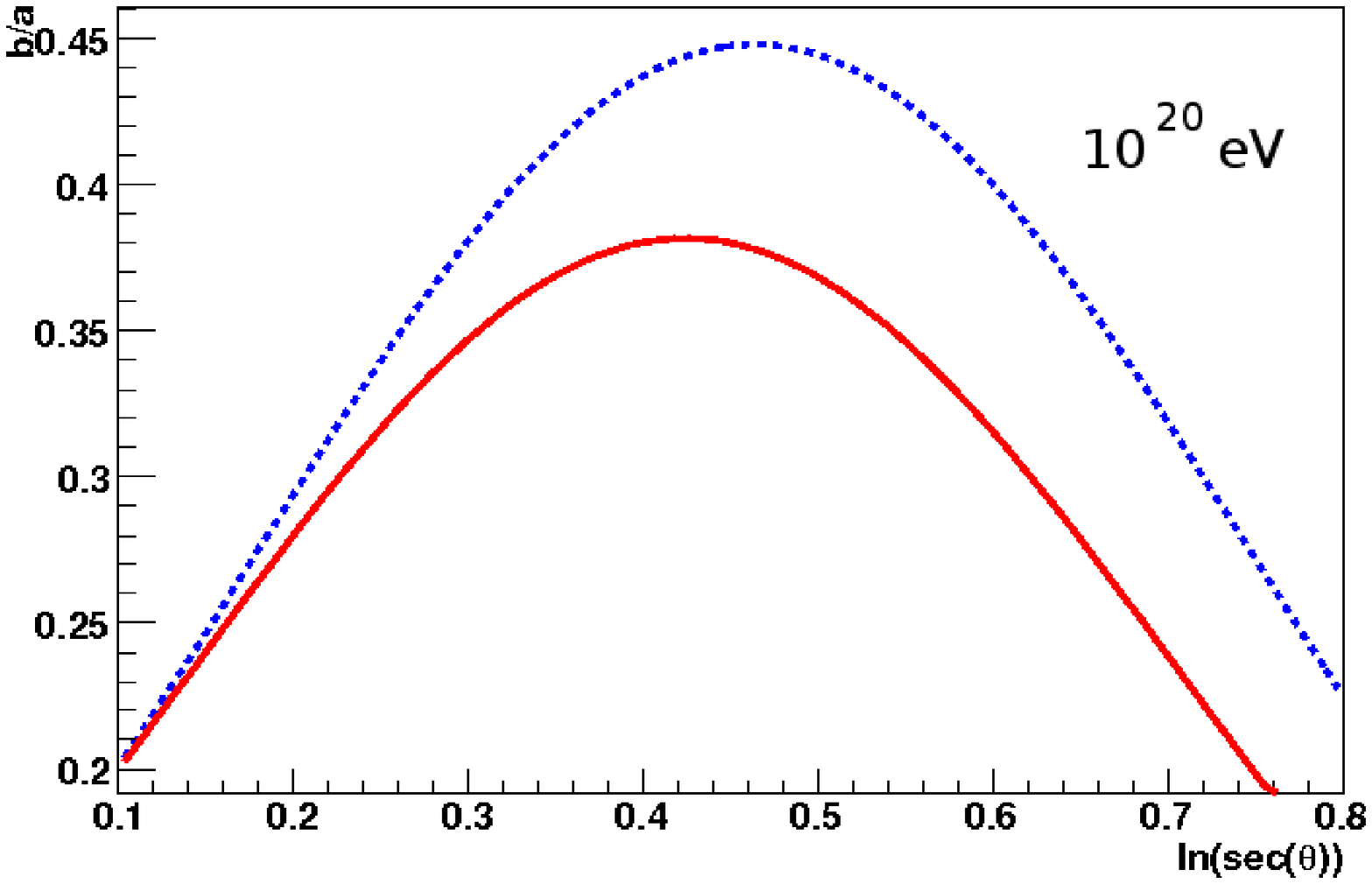}
\caption{Asymmetry longitudinal development at primary energies $10^{18.5}$, $10^{19}$, $10^{19.25}$, $10^{19.5}$, $10^{19.75}$ and $10^{20}$ {\rm eV} (top to bottom from left to right ). Monte Carlo proton 
(dashed line) and iron (solid line) are shown.}
\label{fig:long_mc_sim}
\end{figure}

The procedure used to get the average longitudinal development of the asymmetry, 
can be summarised in the following steps:
\begin{itemize}
\item
Select events in bins of reconstructed energy and ${\rm sec}\, \theta$
values. For these events the risetime of those stations
passing the corresponding cuts is determined. 
For each interval of $E$, ${\rm sec}\,\theta$ and $\zeta$, 
the mean value and the standard deviation 
of the corresponding $t_{1/2}/r$ distribution, is calculated. 
It is worth mentioning that risetime grows nearly linear with the core distance  (see ~\cite{Karlpeople} and references therein) and 
thus the $t_{1/2}/r$ value is more sensitive for the asymmetry as the whole range 
of $r$ can be used for the analysis.

\item 
For each ($E$, $\theta$) bin, a fit of $<t_{1/2}/r>$ to 
a linear cosine function of $\zeta$ (equation \ref{eq:linearcos}) 
provides the asymmetry factor $b/a$.
Figure \ref{fig:rise_vs_zeta_mc} shows as an example, the dependence of 
$<t_{1/2}/r>$  with azimuthal angle for Monte Carlo samples
for both primaries, proton (left) and iron (right), at E = $10^{19}$ {\rm eV} and 
four zenith angles, 32$^o$, 45$^o$, 53$^o$ and 60$^o$. 
The line represents the result of the fit to equation \ref{eq:linearcos}. 
From these plots it can be seen that the mean value of $<t_{1/2}/r>$ 
decreases with the zenith angle as expected. Besides, the asymmetry 
increases with the zenith angle up to a maximum value 
and then decreases for larger angles.
\item 
For each primary type and energy interval the dependence of the asymmetry factor
on ${\rm sec} \theta$ is studied. 
In all cases the plot $b/a$ versus ${\rm ln}({\rm sec}\,\theta)$, i.e. the asymmetry longitudinal development,
is quite symmetric and shows a clear maximum. The position of the maximum is obtained by fitting
a normal function to the values of $b/a$ in bins of  
${\rm ln}({\rm sec} \theta)$. 
Figure \ref{fig:long_mc_sim} shows as an example the results for both primaries
at several energies. 
\end{itemize}

\begin{figure}[htpb]
\includegraphics[width=7.0cm]{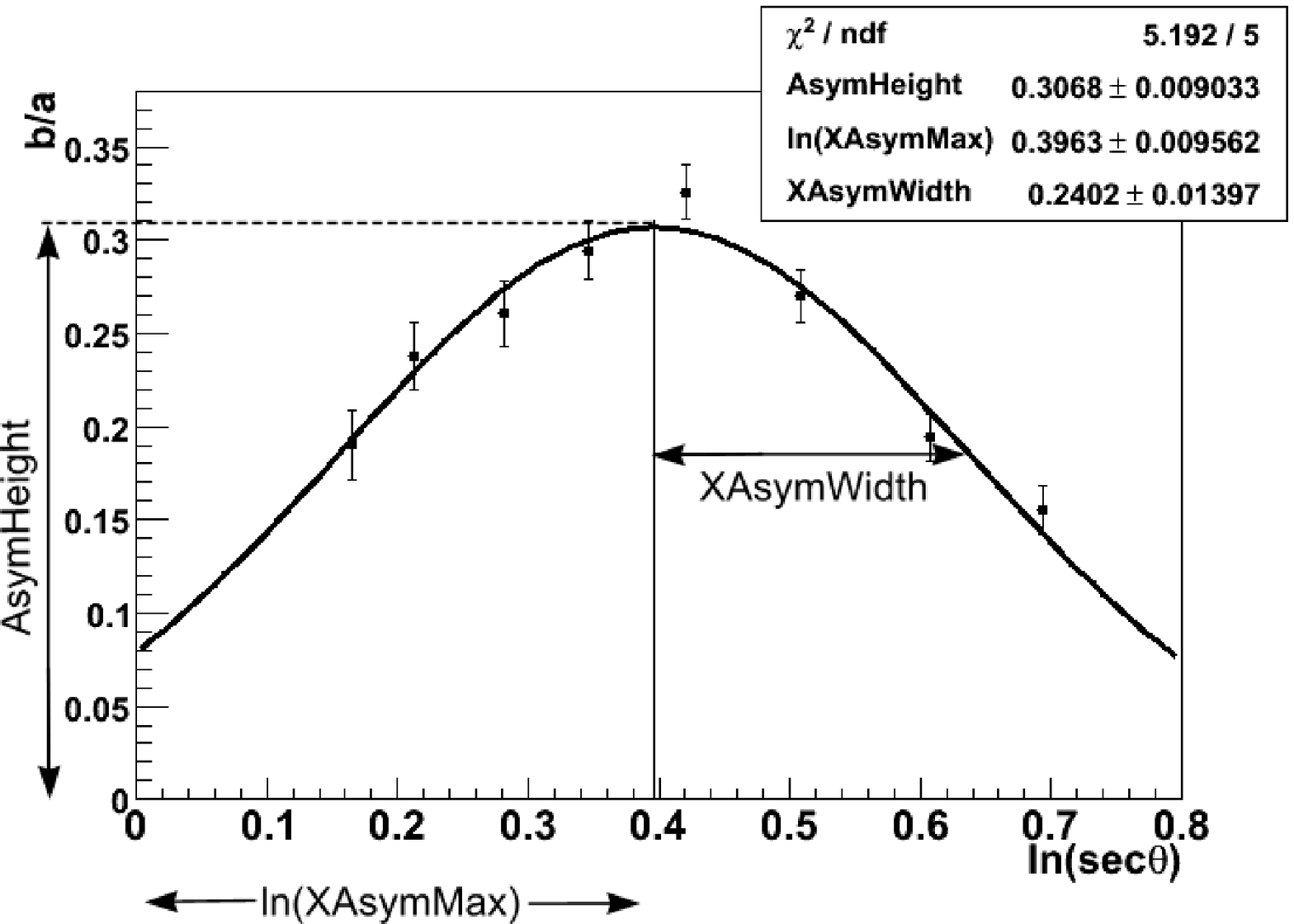}
\includegraphics[width=6.7cm]{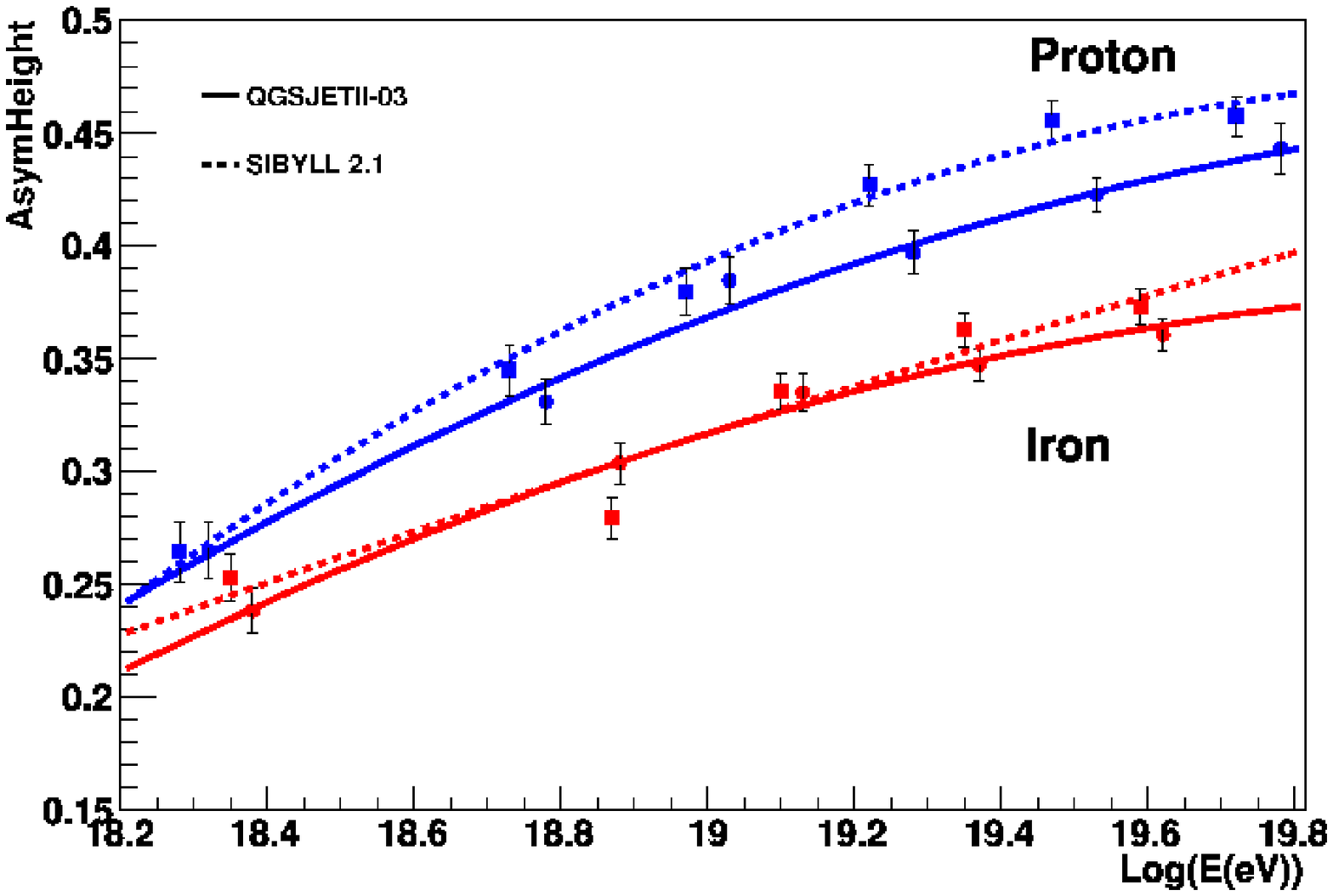}
\includegraphics[width=6.9cm]{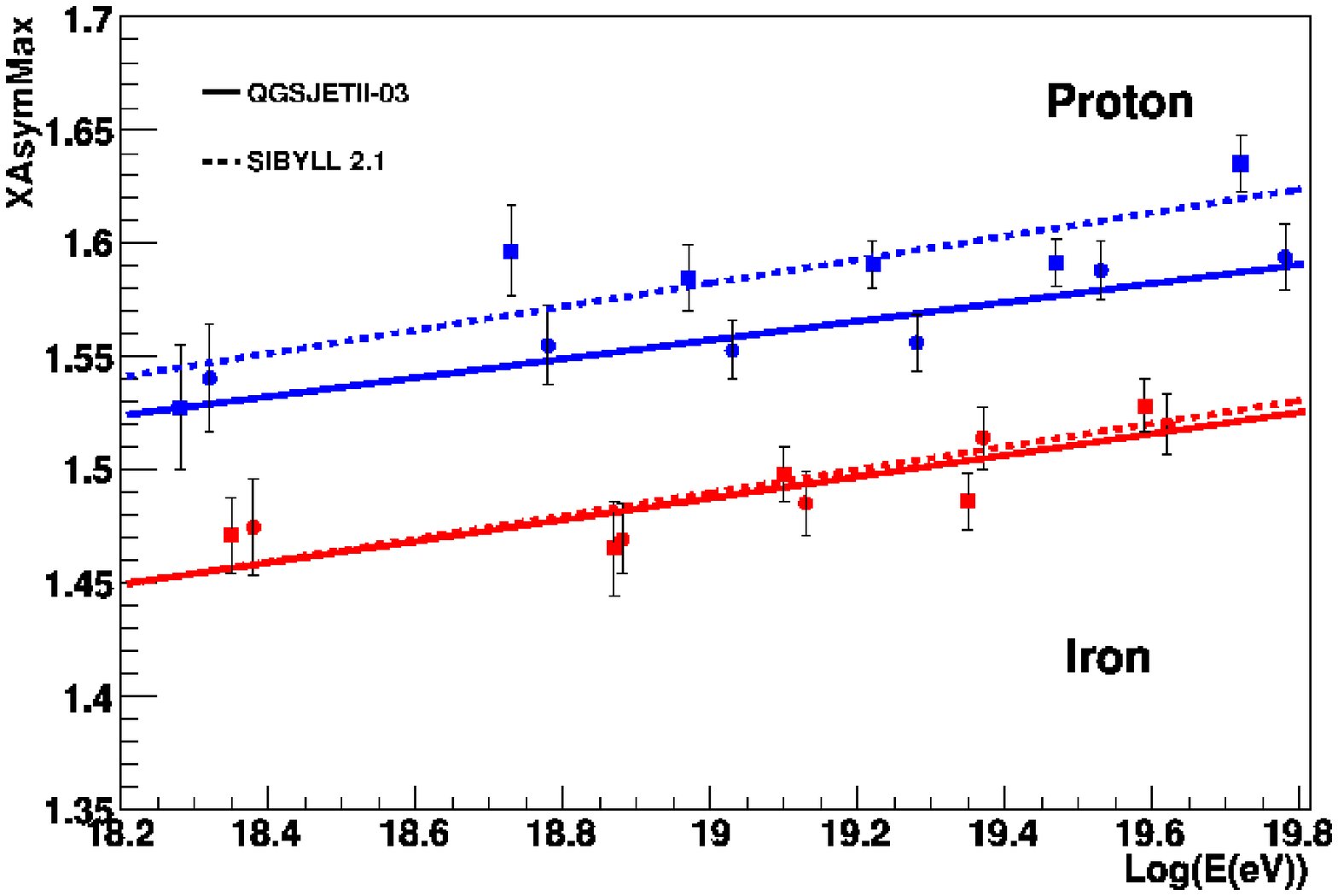}
\includegraphics[width=7.40cm]{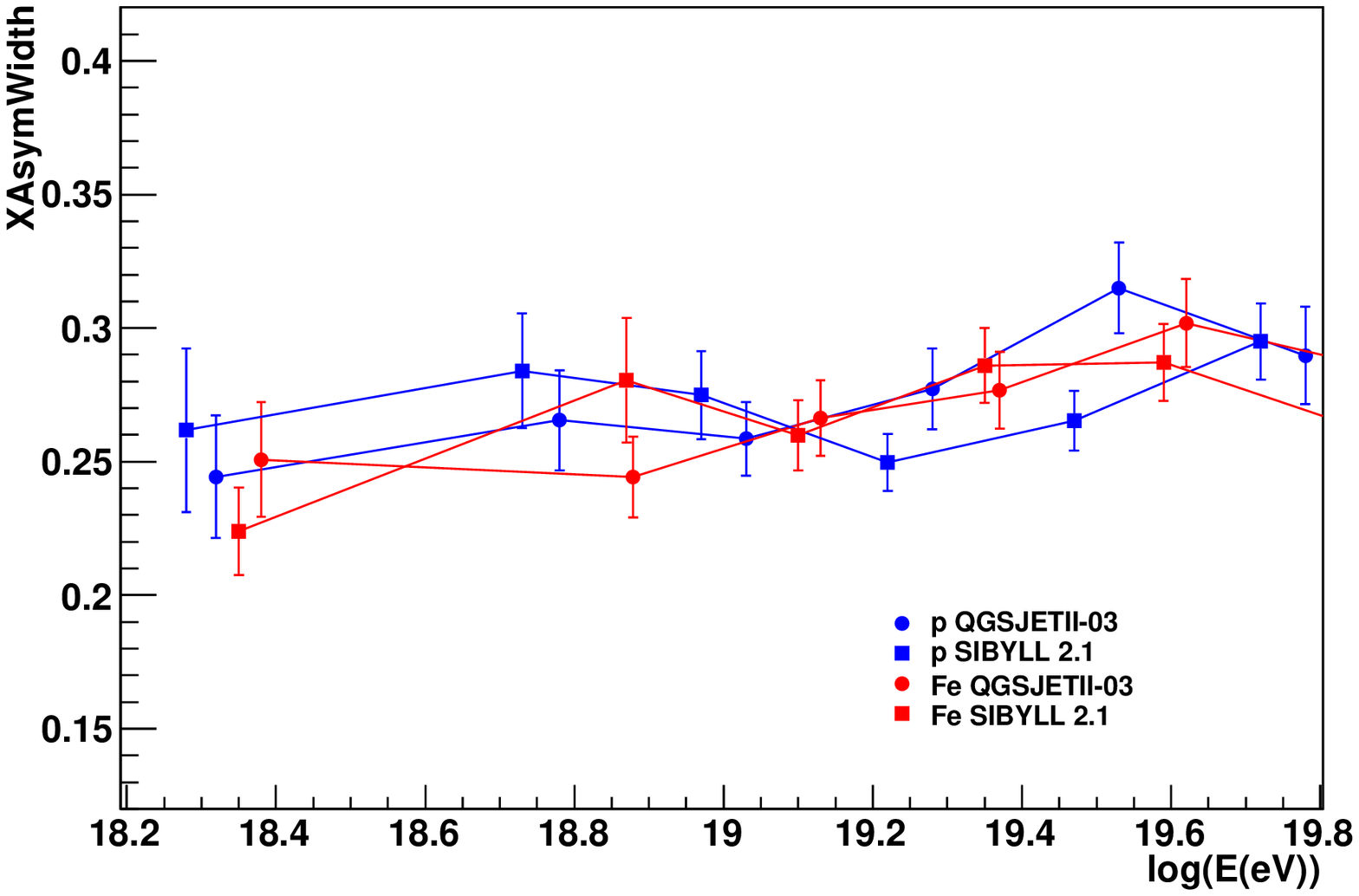}
\caption{Top-left: parameters describing the asymmetry longitudinal development. The other figures show the parameter dependence with primary energy for {\sc qgsjetii}(03) and {\sc sibyll} 2.1. The lines in Figure(bottom-right panel), are just to guide the eye.}
\label{fig:model_had}
\end{figure}

This asymmetry longitudinal development can be described by means of three parameters, 
as indicated in Figure \ref{fig:model_had} (top-left): {\it XAsymMax}, 
the position of the maximum asymmetry, i.e. the ${\rm sec} \theta$ value for which b/a maximises; 
{\it AsymHeight}, the height at maximum, i.e. the maximum b/a value, and {\it XAsymWidth},
 the half width at half maximum of the Gaussian function. 

\subsection{Energy dependence and hadronic models}
\label{sec:had}

In Figure ~\ref{fig:model_had} (top-right panel and bottom panels), the values of the parameters obtained 
from the above mentioned fit have been represented against primary energy. Results for both hadronic models 
are included in the plots. 
The error bars come from the fitting uncertainties. The bottom-left plot shows that {\it XAsymMax} grows 
linearly with {\rm log}$E$. 
The corresponding linear fits (continuous lines) of both primary types are clearly separated, thus allowing 
discrimination of heavy and light primaries. On the other hand, the {\it AsymHeight} parameter 
shows a dependence with log$E$ which nearly follows a parabolic function.
 Results of the corresponding fit are shown for iron (red solid line) and proton (blue solid line) in Figure 
~\ref{fig:model_had} at the top-right panel. The values of {\it AsymHeight} for heavy and light primaries 
for primary energies above $10^{18.5}$ {\rm eV} are well separated. 
Finally, {\it XAsymWidth} is nearly independent on primary energy with a value very similar for both primaries 
and therefore this parameter does not allow separation between primaries. 
The results seem to indicate that the correlation of the asymmetry parameters with $X_{\rm{max}}$  
has a slight dependence on the hadronic interaction model. This will be addressed in the next section. 
As can be seen in Figure \ref{fig:model_had}, predictions for iron do not show a remarkable 
dependence on the model. On the contrary, there is a clear difference for proton primaries, 
particularly at high energies. This is also a feature of the $X_{\rm{max}}$ versus $E$ plots 
(elongation rate) where a similar behaviour at high energies is observed for both models ~\cite{ER:Unger}.

\section{Asymmetry parameters and depth of shower maximum}
\label{sec:xmax}

It was mentioned before that $X_{\rm{max}}$ is the main observable related to composition 
in fluorescence measurements. 
Thus, it is desirable to study the correlation between our asymmetry mass sensitive parameters 
measured with a surface detector and the position of shower maximum. 
To this end, the steps described above were repeated but instead of 
grouping separately p and Fe events by primary energy, they were grouped 
in bins of $X_{\rm{max}}$ mixing both primaries. 
The $X_{\rm{max}}$ values used in these plots, are those of the simulated EAS.

\begin{figure}[htp]
 \begin{minipage}[c]{0.5
\textwidth}
 \centering
\includegraphics[totalheight=0.6\textwidth,angle=0]{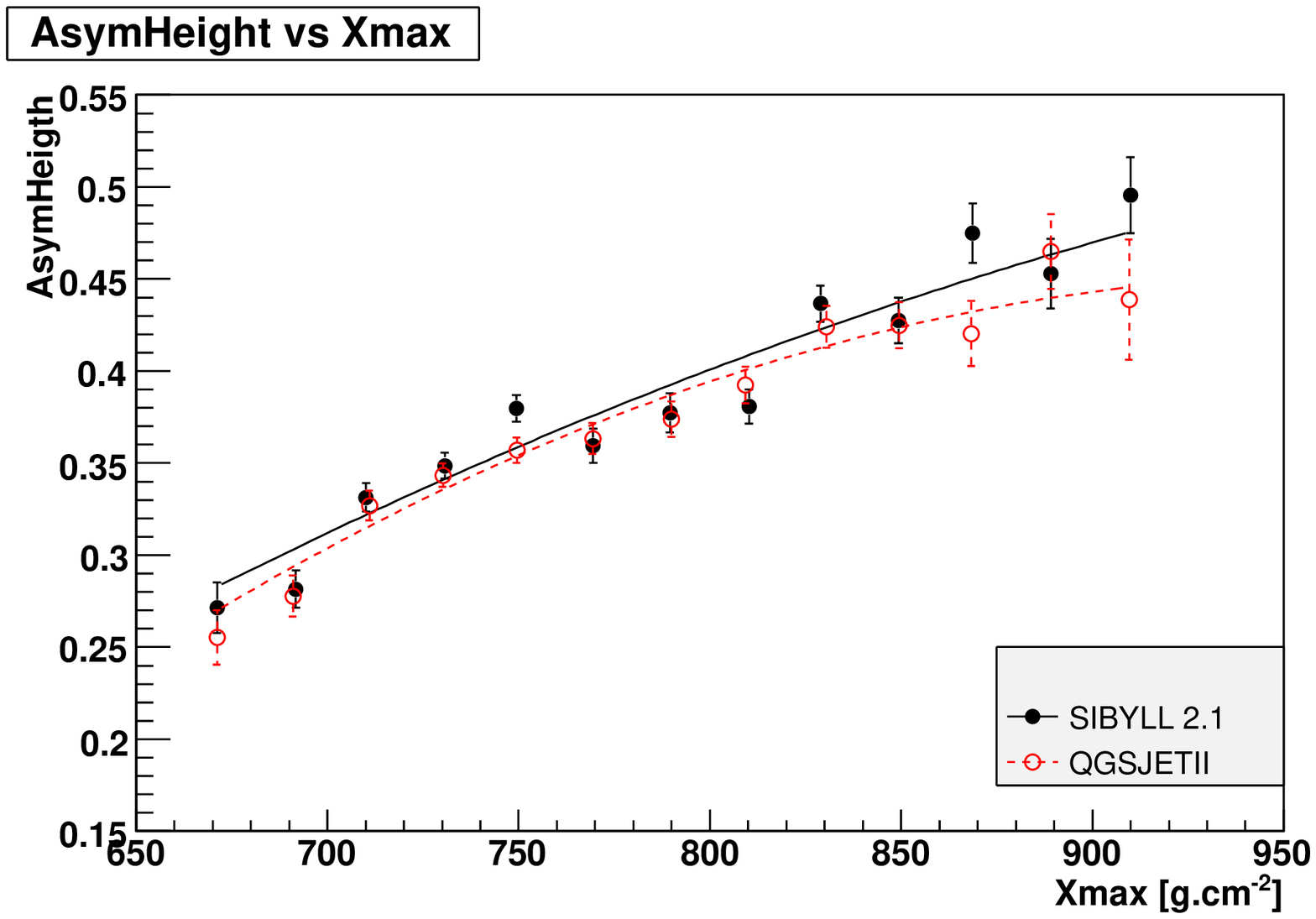}
\includegraphics[totalheight=0.6\textwidth,angle=0]{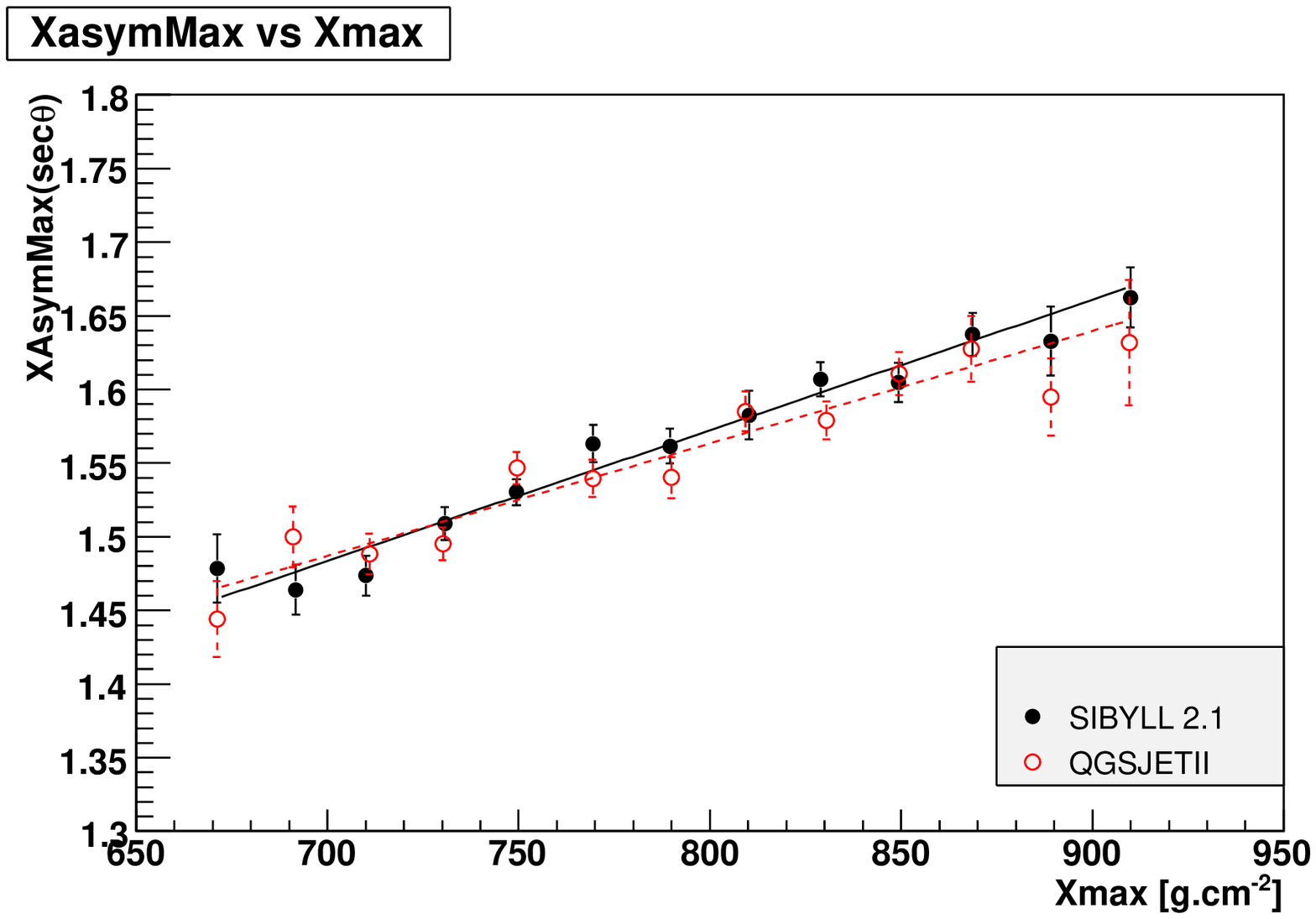}
 \end{minipage}
 \hfill
 \begin{minipage}[c]{0.5\textwidth}
 \centering
\includegraphics[totalheight=0.6\textwidth,angle=0]{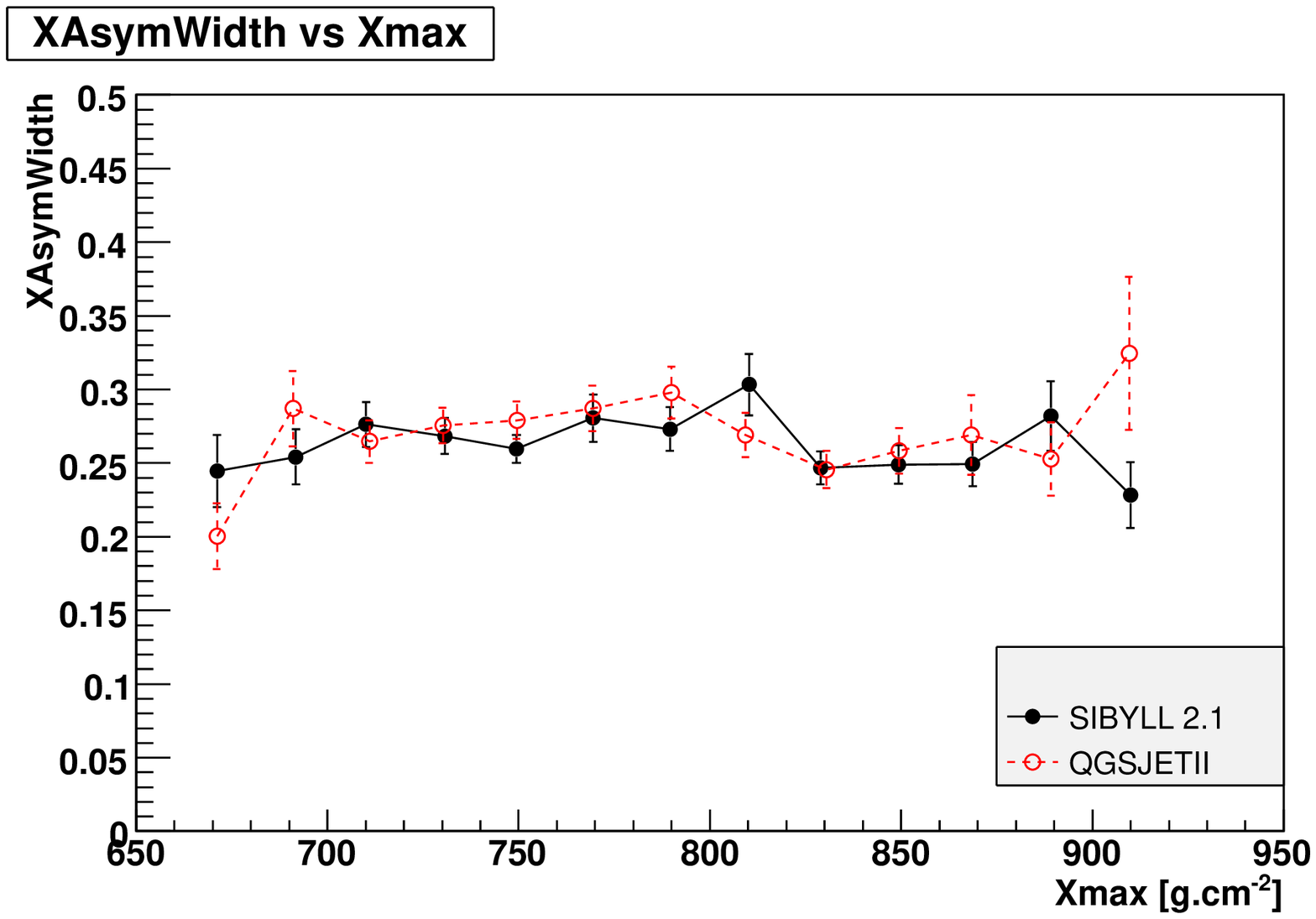}
 \parbox[t]{0.5\textwidth}{\caption{Parameters describing the asymmetry longitudinal development. Correlation with $X_{\rm{max}}$ for {\sc qgsjetii}(03) and {\sc sibyll}}}
\end{minipage}
\label{fig:model_xmax}
\end{figure}

In Figure \ref{fig:model_xmax} the correlation with $X_{\rm{max}}$ is shown for the three parameters: 
{\it XAsymMax}, {\it AsymHeight} and {\it XAsymWidth} 
using both hadronic interaction models {\sc qgsjetii}(03) and {\sc sibyll} 2.1.

As expected from Figure \ref{fig:model_had}, {\it XAsymMax} and {\it AsymHeight} 
present a strong correlation with $X_{\rm{max}}$ which is nearly independent on the 
hadronic model. This is an encouraging result 
reaffirming that the observed azimuthal asymmetry is a reliable mass estimator, providing accurate models to describe EAS.
Certainly, the correlation of the SD parameters with the position of shower maximum 
might be also useful to provide a measurement of  $<X_{\rm{max}}>$ using only the data of surface detector.

\section{Extracting the mean primary mass}
\label{sec:fefraction}

\begin{figure}[htpb]
\includegraphics[width=6.5cm]{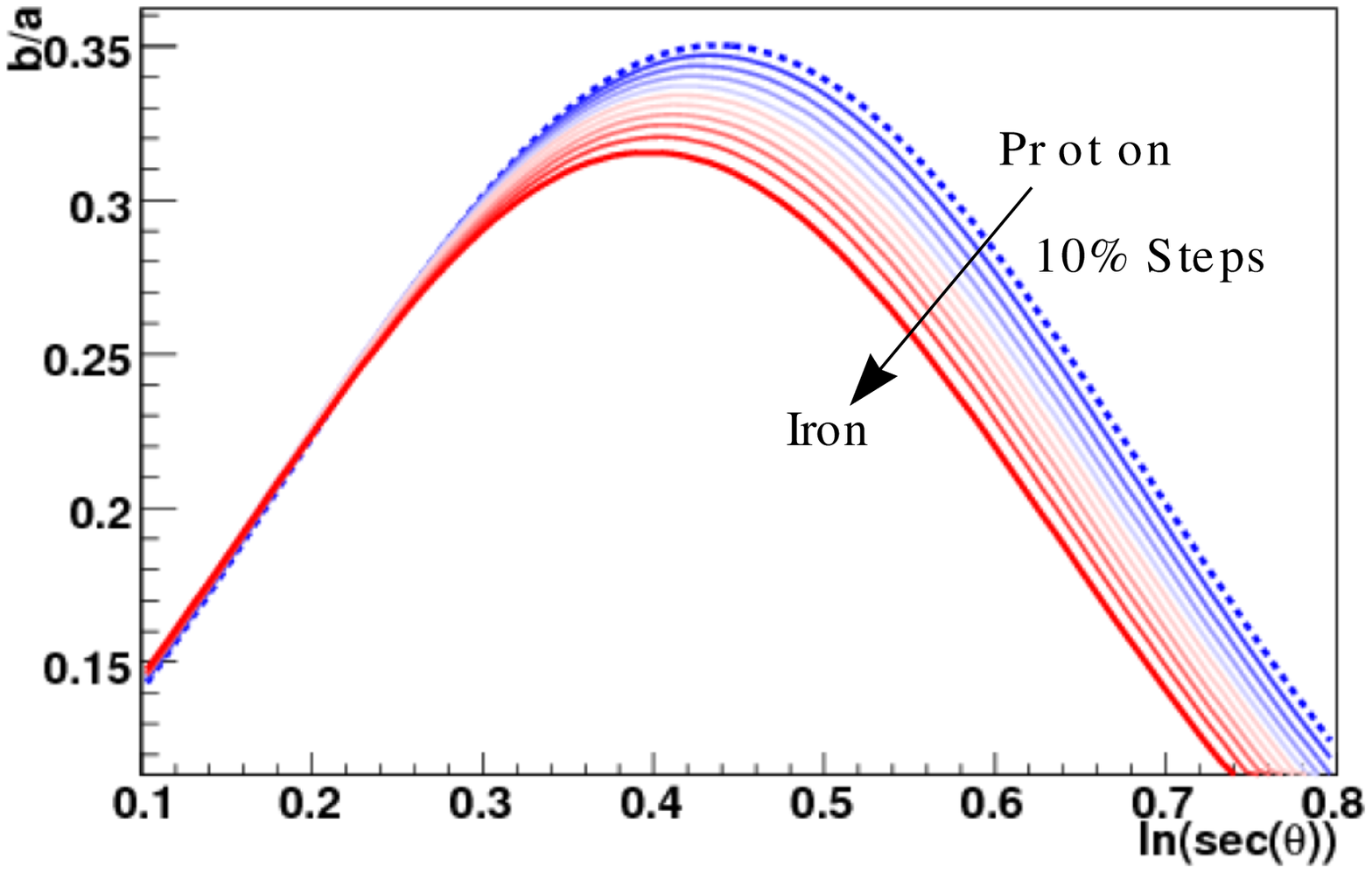}
\includegraphics[width=7.cm]{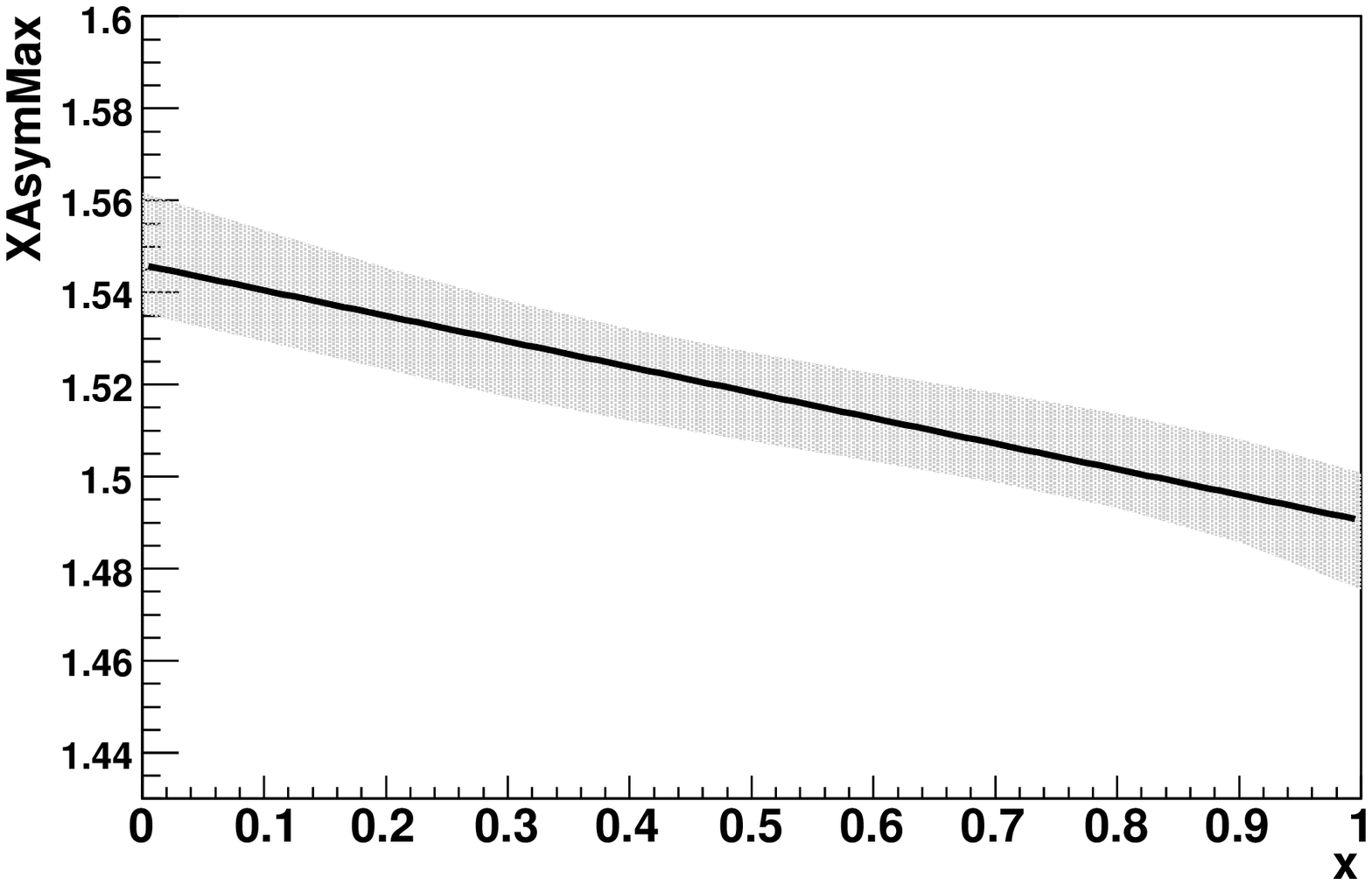}
\includegraphics[width=7.cm]{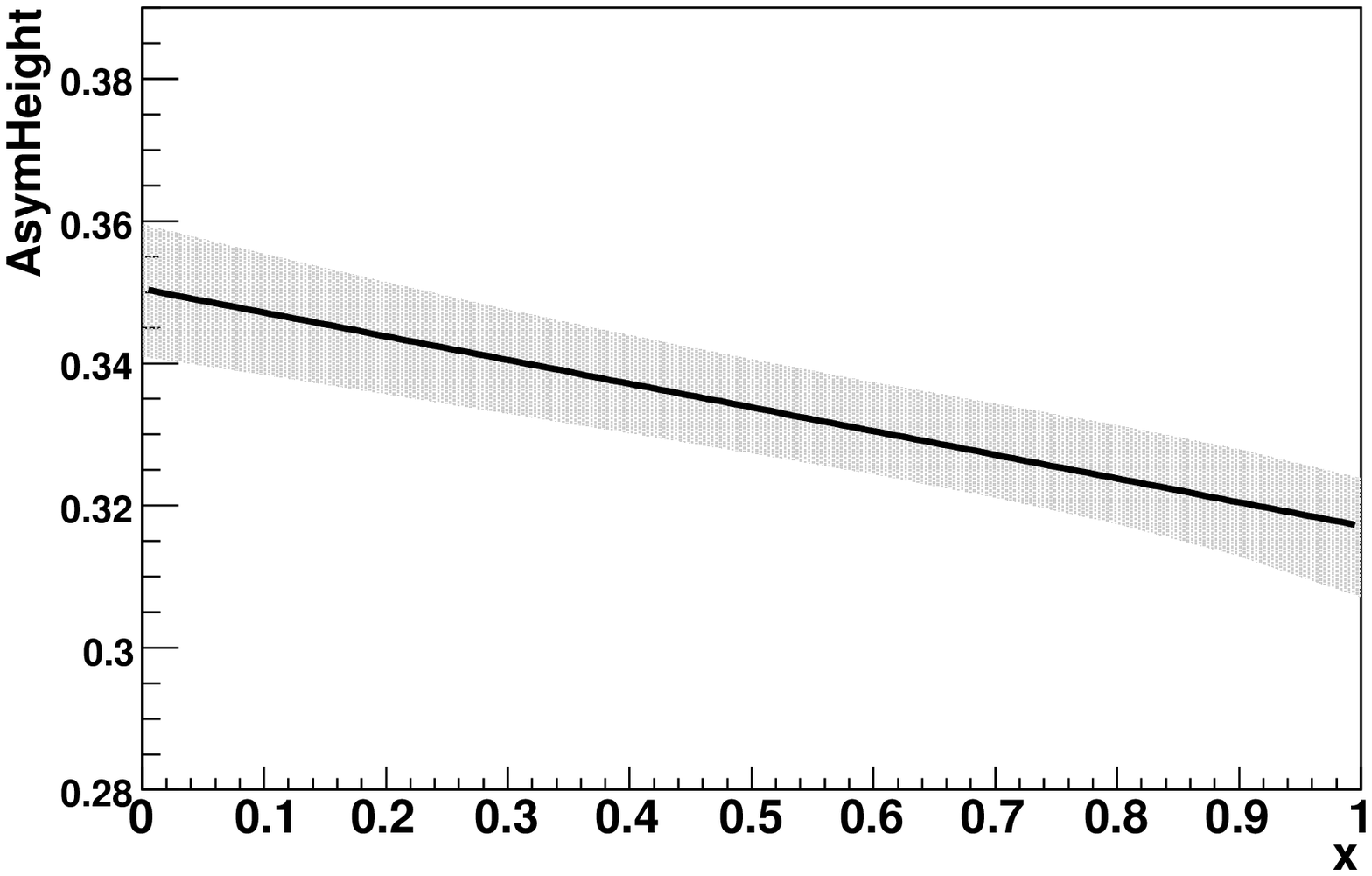}
\includegraphics[width=7.cm]{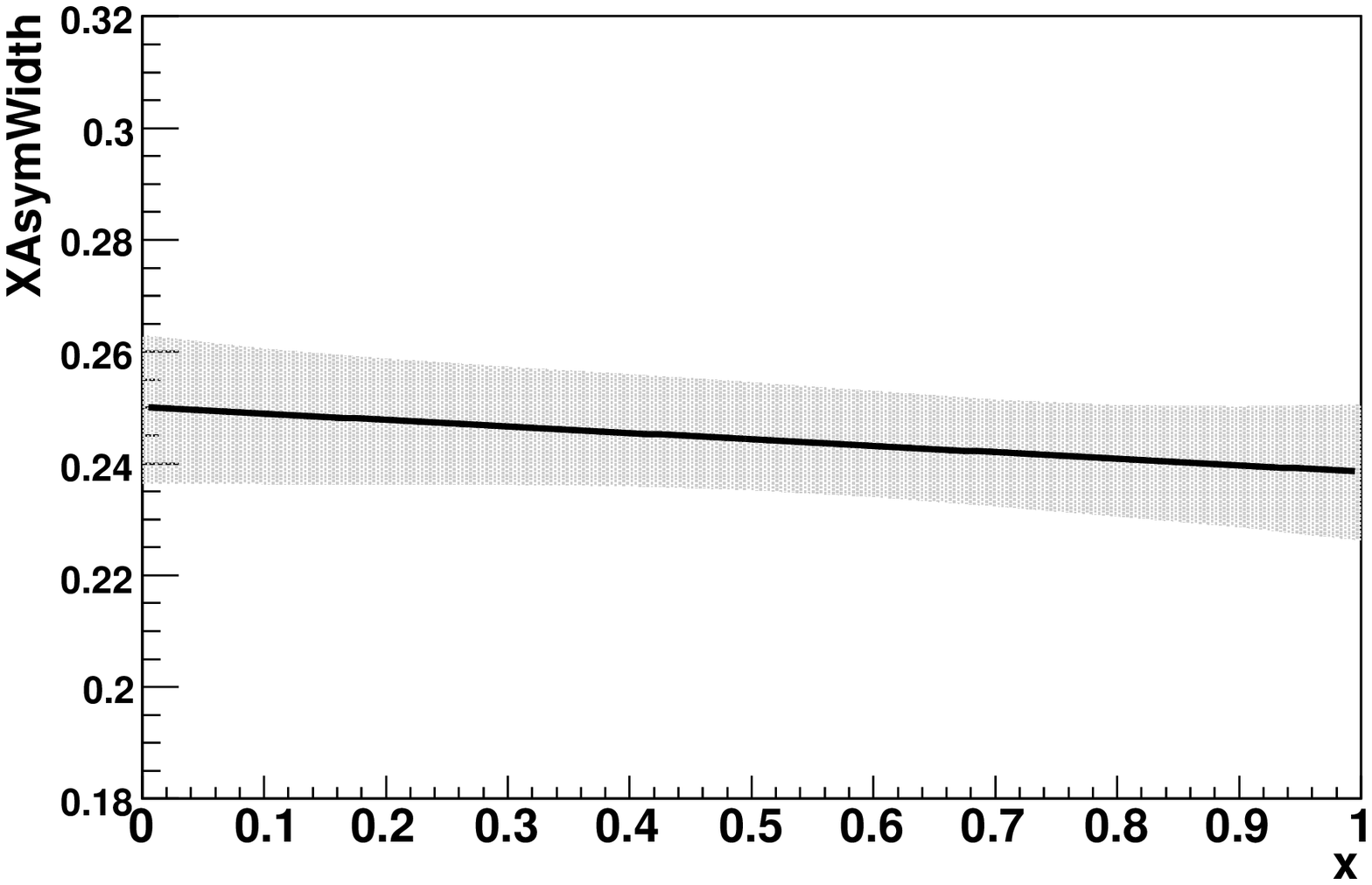}
\caption{Asymmetry development for the different samples with mixed composition, going from pure proton to pure iron in steps of 10$\%$. Position of the maximum, sigma and height of the longitudinal profile as a function of the mixture fraction. 1 corresponds to 100 $\%$ iron and 0 to 100 $\%$ proton. Parameters shown here as an example correspond to primary energy of $10^{19}$ {\rm eV}.}
\label{fig:param_mix}
\end{figure}

As already shown, the asymmetry longitudinal development, 
described by the three parameters above defined, is sensitive 
to the mass composition. The dependence of these parameters 
on composition has been determined as follows. Firstly, a two 
component composition (proton-iron) has been assumed. 
For each interval in $E$, ${\rm sec}\, \theta$ and $\zeta$, the risetime 
for proton and iron showers has been obtained following the 
procedure described in section \ref{sec:steps}. 
The corresponding risetimes for a number of mixtures have been 
inferred as the average value for both components weighted with 
the corresponding composition factors. The percentage of iron and 
proton events in the sample was changed in steps of 10\%. 
As an example, the asymmetry longitudinal development curves 
at $10^{19}$ {\rm eV} showers for different compositions are shown in 
Figure \ref{fig:param_mix} (top-left). A smooth transition 
between pure proton to pure iron compositions can be observed.
This procedure allows us to determine the dependence of 
the three discrimination parameters with composition, e.g. the Fe fraction, 
named in the following $x_{Fe}$. The result for $10^{19}$ {\rm eV} is shown
in Figure \ref{fig:param_mix}. These plots show that {\it XAsymMax} and {\it AsymHeight} 
are strongly correlated with $x_{Fe}$ while the correlation is much weaker 
for {\it XAsymWidth}, as expected from the results of section \ref{sec:steps}. 
For all energy intervals the three parameters follow a linear behaviour, 
decreasing with the iron fraction. The grey area in Figure \ref{fig:param_mix} 
represents the statistical uncertainties, 
determined by the errors bars from the previous fits 
(see Figures \ref{fig:rise_vs_zeta_mc}  and \ref{fig:model_had}). 
The result for a linear fit of these three separation parameters 
is shown in this figure (solid line).
The parameterisation of this dependence will be used later to extract the primary mass.
The composition of a sample, i.e. the $x_{Fe}$ value (either a Monte Carlo test sample or real data) 
can be determined by minimising the following  function 
\begin{figure}[htpb]
\includegraphics[width=14cm]{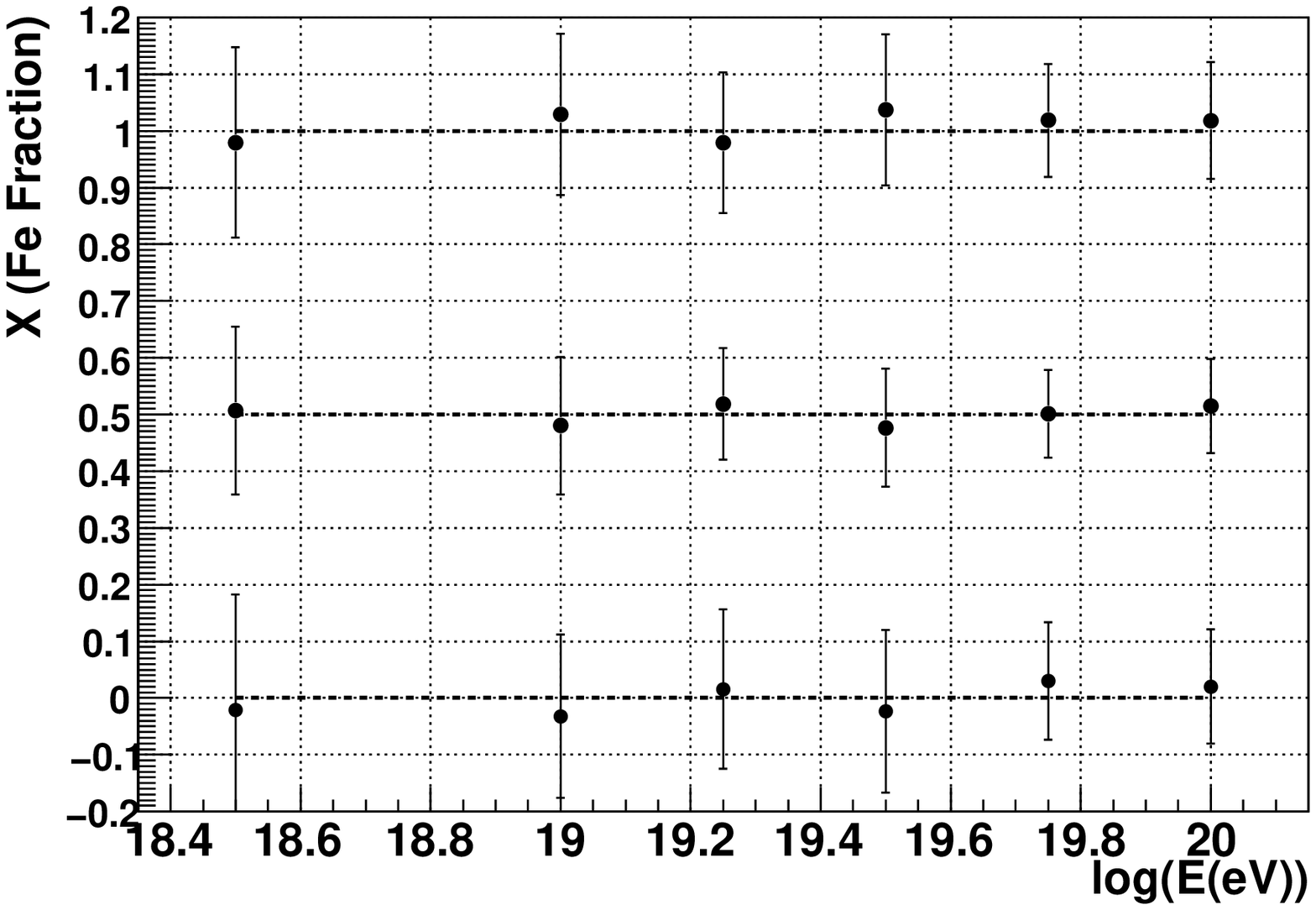}
\caption{Consistency check, iron fraction with three different samples: 100\% iron(upper curve), 50\% iron and 50\% proton(middle curve), 100\% proton(lower curve)}
\label{fig:consistency}
\end{figure}
\begin{equation}
\Delta^2 = \frac{(\sigma_{s} - f_{\sigma}(x_{Fe}))^2}{\Delta \sigma_{s}^2
+ \Delta f_{\sigma}^2} + \frac{(h_{s} - f_h(x_{Fe}))^2}{\Delta h_{s}^2 +
\Delta f_{h}^2} + \frac{(m_{s} - f_{m}(x_{Fe}))^2}{\Delta m_{s}^2 + \Delta f_{m}^2}
\label{eq:delta2}
\end{equation}
where $\sigma$, $h$ and $m$ stand for the parameters {\it XAsymWidth}, {\it AsymHeight}
 and {\it XAsymMax}, respectively; the subindex $s$ indicates the corresponding 
values of the sample to be studied. 
$\Delta \sigma$, $\Delta h$ and $\Delta m$ 
represent the standard deviation of the separation parameters.
The functions $f_{i}(x_{Fe})$ stand for the parameterisation of the dependence of each parameter 
with the iron fraction $x_{Fe}$ obtained in Figure \ref{fig:param_mix}.
The grey area indicates the uncertainty in the parameterisation due to limited Monte Carlo statistics, 
and it will be taken into account as a systematic error. 

To check the reliability of this technique we have reconstructed the $x_{Fe}$ value of several 
samples containing a known fraction of proton and iron showers from our Monte Carlo data by 
minimising $\Delta^2$. 
This test has been carried out using mixtures of showers containing the same events used previously to get the results on Figure ~\ref{fig:param_mix}. The results are shown in Figure ~\ref{fig:consistency} 
for three different samples: 100\% iron, i.e. $x_{Fe}=1$, (upper curve), 
50\% iron and 50\% proton, i.e. $x_{Fe}=0.5$, (middle curve) 
and 100\% proton, i.e. $x_{Fe}=0$, (lower curve). Deviations of reconstructed $x_{Fe}$ values from the real ones are below 0.04 (i.e. errors in the Fe percentage below 4 units). The error bars 
represent the uncertainty due to our limited statistics and amount about $\Delta x_{Fe}$ = 0.1. This of course can be reduced using larger MC samples. 

A study of systematic errors has been performed.
Firstly, the full analysis has been carried out using both hadronic models, 
using the parameterisations from {\sc qgsjetii}(03) 
and then applying that to {\sc sibyll} 2.1 data set.
The prediction in the $x_{Fe}$ value for both models are different 
by an amount which ranges up to $\Delta x_{Fe}$= 0.14. 

Due to the influence of the models the primary energy reconstructed 
for SD events using Monte Carlo, 
deviates from hybrid reconstructed primary energies. 
The effect of this deviation in our
method has been studied by comparing the composition obtained 
using both the MC $E$ 
value (i.e the input for the Monte Carlo generation) and the 
one reconstructed with 
the Auger analysis software. Differences in $x_{Fe}$ lower than 0.04 
have been found. 

A possible contribution due to the cuts applied in our method 
(see section \ref{sec:mc}) 
has been studied by comparing results with different cuts. 
The corresponding contribution 
results in $\Delta x_{Fe}$ = 0.03. 

It should be mentioned at this point that taking the estimated 
event rate of the Pierre Auger Observatory, a similar resolution than the one presented 
in Figure ~\ref{fig:consistency} 
can be achieved at present for energies below $10^{19}$ {\rm eV}. 
With increasing statistics collected by the full array in the near future, the shown resolution will be achieved at higher energies.

\section{Conclusions}
\label{sec:conc}
A novel method to determine the mass composition of primary cosmic rays has been developed using the azimuthal asymmetry in arrival time distribution of secondary particles at a given observing level. The approach relies on statistical grounds and thus provides a mean mass composition of a set of showers at a given energy.

The main idea behind the method is to reconstruct a longitudinal development of the observed  asymmetry which is reminiscent of the longitudinal development of the extensive air shower. A detailed analysis using the risetime of the signal in water Cherenkov detectors for the case of the Pierre Auger Observatory was presented. It was shown that both the atmospheric depth corresponding to the position of maximum asymmetry and the value of the maximum asymmetry are sensitive to primary mass. These parameters measured by the surface detectors were shown to correlate with the position of shower maximum, $X_{\rm{max}}$. 

The method was validated using hypothetical data samples corresponding to pure proton, pure iron and a mixed composition. Systematic uncertainties affecting the determination of primary composition were investigated. As expected, the dominant source of uncertainties comes from the lack of knowledge of hadronic interaction models, which amounts to $\le$ 14\% out of a total of 18\% in the estimation of Fe fraction. The analysis indicates that at the event rate collected by the Pierre Auger Observatory, a very good separation of heavy and light elements can be achieved with present data, for energies below $10^{19}$ {\rm eV}, while at higher energies at least one order of magnitude more data would be needed to achieve the shown resolution in this paper.

\section{Acknowledgements}
The analysis presented here is based on many Auger Offline packages tools, 
for this, we would like to thank the Offline team for providing them.
We gratefully acknowledge the usage of the IN2P3 computing centre in 
Lyon (http://cc.in2p3.fr), where we have generated the shower library for this paper.
We are grateful to the Pierre Auger Observatory collaborators, very specially
to Professors Alan Watson and Luis Epele, for fruitful discussions.
We would like to thank the HELEN Fellowship program for giving us 
the possibility of scientific stays within the collaborating 
institutions hosting the authors of this work.
F. Arqueros and D. Garcia Pinto acknowledge the support of the Spanish MEC
(FPA2006-12184-C02-01), CONSOLIDER - INGENIO2010 program CPAN
(CSD2007-000042) and Comunidad de Madrid (Ref.: 910600). D. Garcia Pinto
acknowledges a PhD grant from Universidad Complutense de Madrid.

\end{document}